\documentclass[fleqn,usenatbib]{mnras}

\usepackage{newtxtext,newtxmath}

\usepackage[T1]{fontenc}

\DeclareRobustCommand{\VAN}[3]{#2}
\let\VANthebibliography\thebibliography
\def\thebibliography{\DeclareRobustCommand{\VAN}[3]{##3}\VANthebibliography}

\usepackage{graphicx}	
\usepackage{float}    
\usepackage{amsmath}	
\usepackage{placeins}	
\usepackage{hyperref}
\usepackage[normalem]{ulem}
\usepackage{makecell}    

\newcommand{\mmp}{$\pm$}
\SetMathAlphabet{\mathbf}{normal}{OML}{mdput}{b}{n}
\newcommand{\mum}{$\mathrm{\mu}$m~}
\newcommand{\odeg}{$^{\circ}~$}

\title[Optical properties of CoPhyLab's dust mixtures]{Spectro-photometric properties of CoPhyLab's dust mixtures}

\author[C. Feller et al.]{
 \parbox{\textwidth}{
 C. Feller$^{1,}$\thanks{E-mail: clement.feller@obspm.fr},
 A. Pommerol$^{1}$,
 A. Lethuillier$^{2}$,
 N. H\"anni$^{1}$, 
 S. Schürch$^{3}$, 
 C. Bühr$^{3}$, 
 B. Gundlach $^{2}$, 
 B. Haenni$^{4}$,
 N. Jäggi$^{1}$, 
 M. Kaminek$^{4}$ and 
 the CoPhyLab Team}
\\~\\
\parbox{\textwidth}{
$^{1}$Physikalisches Institut, University of Bern, Sidlerstrasse 5, CH-3012 Bern, Switzerland;
$^{2}$Institut für Geophysik und extraterrestrische Physik (IGeP), TU Braunschweig, Mendelssohnstr. 3, 38106 Braunschweig, Germany;
$^{3}$Departement für Chemie, Biochemie und Pharmazie, University of Bern, Freiestrasse 3, CH-3012 Bern, Switzerland;
$^{4}$Institut für Anatomie, University of Bern, Baltzerstrasse 2, CH-3012 Bern, Switzerland;
}}

\date{Accepted XXX. Received YYY; in original form ZZZ}

\pubyear{2021}

\begin{document}
\label{firstpage}
\pagerange{\pageref{firstpage}--\pageref{lastpage}}
\maketitle

\begin{abstract}
 Objective: In the framework of the Cometary Physics Laboratory (CoPhyLab) 
 and its sublimation experiments of cometary surface analogues under 
 simulated space conditions, we characterize the properties of intimate 
 mixtures of juniper charcoal and SiO$_2$ chosen as a dust analogue 
 \citep{Lethuillier_2022}. We present the details of these investigations 
 for the spectrophotometric properties of the samples.\\
 Methods: We measured these properties using a hyperspectral imager and a 
 radio-goniometer. From the samples' spectra, we evaluated reflectance 
 ratios and spectral slopes. From the measured phase curves, we inverted 
 a photometric model for all samples. Complementary characterizations were 
 obtained using a pycnometer, a scanning electron  microscope and an 
 organic elemental analyser.\\
 Results: We report the first values for the apparent porosity, elemental 
 composition, and VIS-NIR spectrophotometric properties for juniper 
 charcoal, as well as for intimate mixtures of this charcoal with the 
 SiO$_2$. We find that the juniper charcoal drives the spectro-photometric 
 properties of the intimate mixtures and that its strong absorbance is 
 consistent with its elemental composition. We find that SiO$_2$ particles 
 form large and compact agglomerates in every mixture imaged with the 
 electron microscope, and its spectrophotometric properties are affected 
 by such features and their particle-size distribution. We compare our 
 results to the current literature on comets and other small Solar System 
 bodies and find that most of the characterized properties of the dust 
 analogue are comparable to some extent with the spacecraft-visited 
 cometary nucleii, as well as to Centaurs, Trojans and the bluest TNOs. 
\end{abstract}

\begin{keywords}
techniques: photometric -- techniques: spectroscopic -- methods: data analysis -- comets: general
\end{keywords}



\section{Introduction}
Small bodies of the Solar System are tracers of the evolutionary 
processes that occurred since the earliest stages of its formation 
from the Solar nebula. Characterising not only these objects' 
dynamical and bulk properties, but also their surface properties 
is an essential step to complete our understanding of their nature, 
and thus to refine our knowledge of the Solar System's history. 
While an handful of sample return missions have granted a 
first-hand knowledge of the compositional and surface properties 
of a few small bodies, ground- and space-based remote sensing have 
provided the bulk of our understanding for asteroids, comets and 
other small bodies of the Solar System.\\

Comets, in particular, are virtually time-capsules, and the study of 
their physical properties as well as their composition, largely 
preserved since their formation more than 4.5 billions of years ago, 
has allowed to make direct inferences on the properties of the 
proto-planetary disk from which they accreted. \citep{Davidsson_2016, 
Rubin_2019, Drozdovskaya_2019}. Over the last forty years, successive 
space missions have allowed us to observe and investigate comae and 
nuclei in ever greater details (e.g. \citealt{Muench_1986, 
AHearn_2011}). To this day, the ESA-led \textit{Rosetta} mission 
provided the most complete picture of a comet, as it accompanied 
67P/Churyumov-Gerasimenko (hereafter 67P/C-G) throughout its 2015 
perihelion passage. From August 2014 to September 2016, the Rosetta 
spacecraft provided a multi-instrument monitoring of the comet's inner 
coma and its nucleus' surface properties, and in November 2014, the 
Philae probe performed in-situ measurements as it landed on the 
nucleus \citep{Taylor_2017}.\\
The chemical compounds directly identified in the inner coma and at 
the surface of 67P/C-G's nucleus \citep{Altwegg_2019, Haenni_2022, 
Bardyn_2017, Krueger_2017} complete our understanding of cometary 
composition derived, from instance, from the laboratory analysed dust 
particles of comet 81P/Wild-2's coma \citep{Brownlee_2012, 
Sandford_2021}. The quantitative interpretation of remote-sensing data 
suffers however from the lack of ground-truth. While models exist that 
relate photometric observables to physico-chemical properties of the 
surfaces, some of the approximations are purely empirical and the 
complexity of natural surfaces often results in degeneracies when 
inverting the models. In this context, laboratory studies of the 
properties and behaviour of surface analogues under cometary conditions 
can help to provide a more thorough understanding and interpretation of 
remote-sensing measurements to better refine scenarios for the formation 
and evolution of comets.\\

The Cometary Physics Laboratory (CoPhyLab) project follows a 
multi-disciplinary approach to characterize both the evolution of the 
physical properties and the phenomena on and inside the surface of a 
cometary analogue under simulated space conditions, using multiple 
instruments \citep{Kreuzig_2021}. One of the initial objectives of the 
CoPhyLab project consisted in the definition of a suitable and workable 
cometary analogue recipe as a compromise between the suitability of the 
chosen materials, their level of characterization as well as practical 
considerations such as ease of procurement, cost and relative harmlessness 
\citep{Lethuillier_2022}. Of particular importance for many investigations 
planned in this project is the albedo of the final dust simulant as this 
property has a direct impact on the absorption of solar radiation and 
therefore on the thermo-physical processes that affect the material. 
Other photometric properties should ideally be as close as possible to 
actual known properties of cometary nuclei to ensure that the light 
scattering regimes are comparable. This facilitates the direct comparison 
between experimental results and remote-sensing datasets and the 
validation on laboratory samples of the physical models ultimately used 
to invert composition and physical properties from reflectance. Studying 
chemical processes was not an objective of the CoPhyLab project and 
therefore reproducing the known chemical composition of the nucleus was 
not a primary criterion in the analogue selection. As the composition 
of the analogue differs significantly and is much simpler than the one 
of actual cometary material, one cannot expect all spectro-photometric 
properties to match. It is nevertheless important to characterize 
their differences to interpret correctly the implications of laboratory 
results in the context of actual comets observations.\\

Past attempts at simulating cometary processes in the laboratory have 
made use of a variety of materials. During the KOSI experiments, olivine, 
kaoline and montomorillonite were intimately mixed with carbon to produce 
a dark cometary dust simulant, which was then mixed with ice. 
\citet{Oehler1991} note however that the addition of carbon in the mixture 
masks all absorption features by minerals, flattening the spectrum. Carbon 
and carbon-based compounds have been used in a number of other studies to 
achieve a low albedo \citep{stephens1991, moroz1998}. More recently, 
opaque minerals have also been used for a similar purpose. 
\citep{Quirico_2016, Rousseau_2017}. In the framework of the CoPhyLab 
project, the search for a suitable dust analogue ultimately resulted in 
the selection of silicon dioxide and juniper charcoal as the two components 
for the first dust recipe of the CoPhyLab project. As a complement to the 
companion study of \citealt{Lethuillier_2022}, the present work was 
performed to characterize the spectroscopic and photometric properties of 
the two materials chosen to simulate cometary surfaces, as well as to 
investigate the properties of intimate mixtures formed from these two 
materials. Mixed together in the right proportions, these two components 
display an albedo similar to the nucleus while other key physical and 
spectro-photometric properties are also reasonably close. The measurements 
of our study are a counterpart to those obtained, for instance, by the 
OSIRIS and VIRTIS instruments of the ROSETTA mission. Combined, such 
measurements allow to correlate the reflectance of a material with both 
its compositional and physical properties, e.g. through the identification 
of specific absorption bands, or the estimation of the albedos. Furthermore, 
such measurements will serve as references for future CoPhyLab sublimation 
experiments, and complement other physical measurements described in 
\citealt{Kreuzig_2021}, to provide a more accurate modelling of the analogue 
material and the subsequent interpretation of both CoPhyLab and 
remote-sensing observations.\\

In section 2, we present the materials and the instruments used as well as 
the data reduction process before outlining the photometric model used in 
this study. In section 3, we present the results of the spectroscopic 
measurements and of the photometric modelling. Finally, we discuss our 
results in the light of the existing literature in section 4.
\section{The materials, instruments and photometric model used}
\label{sec:material_instruments}
\subsection{End-members of the CoPhyLab dust recipe}

\begin{figure*}
 \begin{center}
 \begin{minipage}[c]{0.49\linewidth}
   \includegraphics[width=\linewidth]{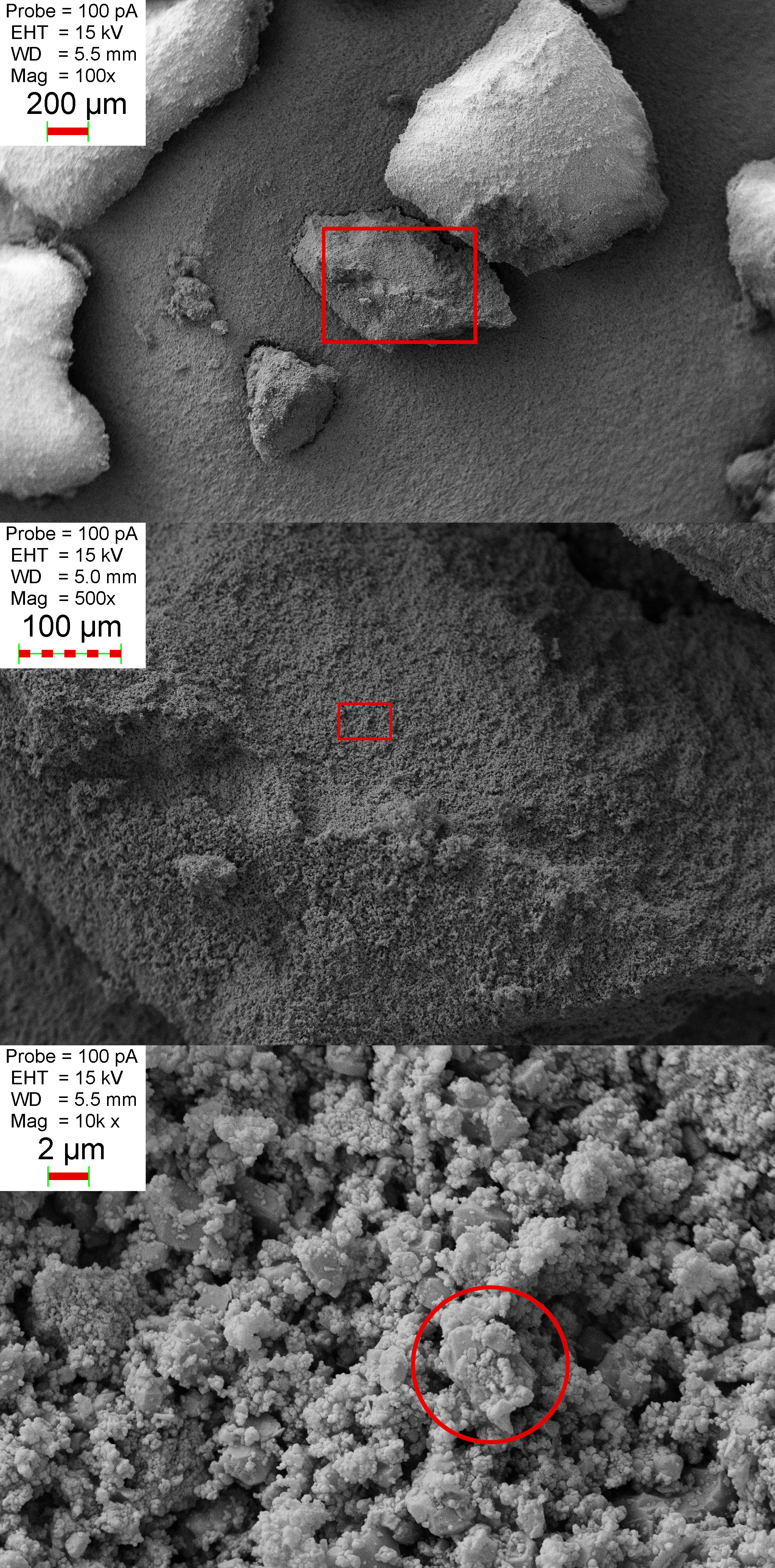}
   \label{fig:sem_sio2}
 \end{minipage}\hfill
 \begin{minipage}[c]{0.49\linewidth}
   \includegraphics[width=\linewidth]{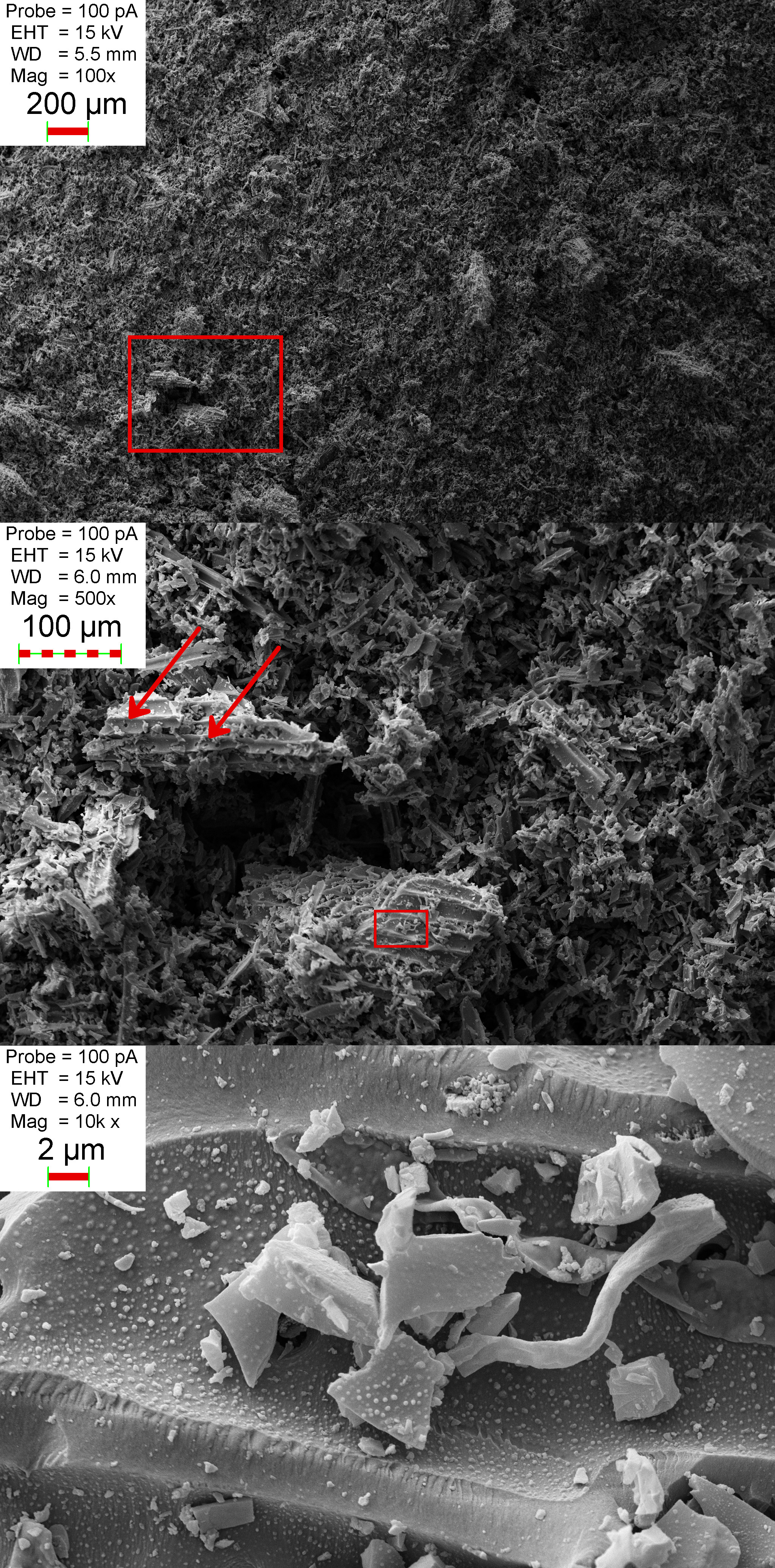}
   \label{fig:sem_jchc}
 \end{minipage}
 \caption{\label{fig:sem_pure} SiO$_{2}$ and JChc samples observed under a 
   scanning electron microscope (SEM) at increasing magnification settings 
   (from top to bottom).
   Left column: Zooming-in on millimetre-sized SiO$_{2}$ aggregates reveals 
   nanometre-sized grains heaping on each other as well as upon larger 
   particles, such as the one encircled in red in the bottom picture. -- 
   Right column: Zooming-in on the juniper charcoal sample exposes a tangle 
   of micrometre- and nanometre-sized wood vessel structures and fragments, 
   such as wall pores (red arrows, middle picture) and lignin protuberances 
   (warty surfaces, bottom picture), which survived the carbonisation, 
   grinding and sieving processes.}
 \end{center}
\end{figure*}

The silicon dioxide powder (SiO$_{2}$, CAS: 14808-60-7) was obtained 
from Honeywell (REF: S5631). Although the particle size distribution 
(PSD) ranges from 0.5 $\mu$m to 10 $\mu$m, as reported in 
\cite{Kothe_2013}, we observe that the SiO$_{2}$ particles easily form 
large aggregates with a diameter up to a few millimetres. Scanning 
electron microscope images (Fig \ref{fig:sem_pure}, left column) 
illustrate that such large SiO$_{2}$ aggregates (top image) are 
formed by the apparent regular assemblage of grains a few hundreds of 
nanometres in size and micrometer-sized chunks (bottom image), 
sticking to and stacked upon each another.\\

As reported in \citet{Lethuillier_2022}, we performed porosity m
easurements through two methods: using a 1 mL graduated flask to 
determine the powder's bulk density, and using an helium pycnometer 
(Upyc-1200e-V5.04) to measure the apparent particle volume (i.e. the 
sum of the solid material volume and that of any closed void), thus 
allowing to retrieve the apparent particle density \cite{Svarovskly_1987}.

Assuming homogeneous particles, we found the SiO$_{2}$ powder sample to 
have a bulk density of 0.885\mmp 0.001 g.cm$^{-3}$, and an apparent 
particle density of 2.59\mmp 0.02 g.cm$^{-3}$. The associated porosity 
for this sample thus stands at 65\mmp 3\%. These values are consistent 
with prior measurements on a similar material using different methods 
\cite{Kothe_2013}.\\

The juniper charcoal powder (hereafter, JChc) was obtained from the 
firm Werth-Metall\footnote{No CAS number is associated with this 
particular substance. Basic information and a picture of the sample 
are accessible at the following address: 
\url{https://werth-metall.de/produkt/wacholderkohle-gemahlen/\#about}, 
last retrieved on \today.}. No specific information could be 
obtained on the manufacturing process, except that the powder had 
been sieved using a 100 \mum mesh sieve. As per the initial CoPhyLab 
dust recipe, the juniper charcoal sample was further dry-grounded 
using an agate disk mill (of the Retsch RS200 model), set at 700 rpm 
for about two and a half minute, and then the smallest particle-size 
fraction was extracted using a 50 \mum mesh sieve.\\
As for the SiO$_{2}$ sample, SEM images of the JChc sample are 
presented in Fig. \ref{fig:sem_pure}, right column). Even at the 
lowest magnification setting (right column, top picture), the JChc 
sample appears to contain a number of elongated (“splinter”-like) 
particles with dimensions reaching $\sim$ 70 \mum emerging out, or 
laying on top, of an agglomerate of smaller particles. Such flat 
splinters can pass in the diagonal of the 50 \mum square mesh sieves. 
Some of these particles (center and bottom images of Fig. 
\ref{fig:sem_pure}, right column, as well as Fig. 
\ref{fig_supp:sem_jchc_psd}) clearly display the structures of wood 
cells (e.g. \citealt{FPL_1980}, \citealt{Jiang_2018}), such as pores 
on vessel walls (pointed by the red arrows in Fig. \ref{fig:sem_pure}), 
or lignin protuberances (the distinct wart-like surfaces in the bottom 
image of the same figure). While features such as pores are $\sim$ 2 
\mum in diameter, lignin protuberances are a few hundreds of nanometre 
in size. As illustrated by the center and bottom images of Figs. 
\ref{fig:sem_jchc} and \ref{fig_supp:sem_jchc_psd}, smaller JChc 
particles exhibit diversity in shapes: from sub-micrometre grains  
through micrometre-sized pieces and platelets to 
tens-of-micrometre-sized fragments of vessel walls.\\

Finally, the bulk and apparent particle densities of the JChc sample 
were found to be respectively equal to 0.479\mmp 0.001 g.cm$^{-3}$ and 
1.5\mmp 0.1 g.cm$^{-3}$, thus giving a porosity of 68\mmp 3\%. 
Although no literature references were found for this specific charcoal, 
these values are consistent with those obtained for different types of 
charcoal, such as beech charcoal \citep{Brocksiepe_2000}.

\subsection{Preparation of the mixtures}
As presented in \citet{Lethuillier_2022}, a series of 9 intimate binary 
mixtures of SiO$_{2}$ and JChc were prepared with steps of 10 wt.\% in 
the JChc mass fraction. The first mixture thus contained 10\% JChc by 
mass, and the last one 90\%.\\
In order to produce these mixtures, we proceeded systematically in 
this way: the whole of the minor fraction was first placed in a bottle 
in which the major fraction was then progressively introduced. With 
each addition of the major fraction material, the bottle was gently 
shaken by hand for 10 seconds to homogenize the mixture. Any agglomerate 
larger than a few millimetres, that formed during the shaking, was 
broken apart with a spatula. We decided to use a manual shaking as 
initial tests with a mechanical shaker immediately produced a number 
of such large agglomerates.\\
Using this procedure, $\sim$ 12 g of each mixture were prepared, 
an amount sufficient to fill either type of sample holder used in the 
radio-goniometric and spectroscopic measurements.\\

For either type of measurements, all of the samples were prepared 
in a similar manner. Black-anodized aluminium containers were filled up 
to rim using a spatula and a final layer of material was sprinkled using 
a large sieve (1.5mm mesh size) to obtain a random surface orientation 
and avoid material compression.\\
The pictures of the samples prepared in this way are presented in Fig. 
\ref{fig_supp:pics}.
\subsection{Spectroscopic measurements}
Using the Mobile Hyperspectral Imaging Setup (MoHIS) of the University of 
Bern, hyperspectral VIS-NIR cubes with a high spectral resolution were 
acquired for samples of each of the end-members and prepared mixtures.\\
The instrument initially designed as a spectro-imaging system for the 
Bernese pressure and temperature simulation chamber called SCITEAS 
\citep{Pommerol_2015}, was revised to become portable and 
convenient for use with other simulation chambers \citep{Pommerol_2019}.\\

The light source of MoHIS consists of a 250 W halogen lamp (Newport/Oriel 
\# 67009) and a gratings monochromator (Newport/Oriel MS257), which allows 
to sweep the near-UV to near-infrared wavelength range with adjustable step 
and bandpass. A optical fibre bundle, coupled with the monochromator, guides 
the light up to a collimating lens, placed at the centre of MoHIS' detector 
head. Hence, when this part is attached to the flange of a simulation 
chamber, the lightbeam's boresight is normal to the flange' surface.\\

The imaging system consists of two cameras, one sensitive in the 
visible domain, the other in the near-infrared domain. The visible 
camera (with a 1392x1040 pixels CCD detector and an image scale of 0.46 
mrad/pixel) images the samples across the 395 nm to 1055 nm wavelength range, 
while the IR camera (with a 320x256 pixels MCT detector and an image 
scale of 2.25 mrad/pixel) images the same surface from 800 nm to 2450 nm.\\
Each camera is pointed towards one of two 45\odeg folding mirrors, which 
are fixed diametrically opposed with respect to the collimating lens, and 
as close to it as possible. Both mirrors are turned towards the photocentre 
of the lightbeam.\\
This setup allows to illuminate a sample placed directly underneath it at 
low incidence and emission angles. Because of the proximity between the 
sample, the cameras and the illumination system (30 to 40 cm), the exact 
values vary slightly across the image and in between the VIS and the NIR 
camera but the phase angle remains in a range of 4 to 7 \odeg.\\

As discussed in \citet{Jost_2017}, the monochromator can be set to sweep 
the wavelength domain with spectral sampling and resolution adapted to 
the width of the spectral signatures investigated. For the experiments 
of this study, we used a spectral sampling of 15 nm in the visible range 
and 6 nm in the infrared range. The spectral resolution (full-width at 
half-maximum of the transmitted band) was 6.5 nm from the visible to 1300 
nm and 13 nm at longer wavelengths in the infrared.\\

Following their acquisition, all the images are first calibrated to 
reflectance factor (REFF), and then assembled into hyperspectral datacubes. 
For the purpose of that calibration, for each sample measurement, 
an additional “reference” cube is also acquired by imaging a 15x15 cm$^2$, 
99\% reflective Spectralon$^{\text{\textregistered}}$ (LabSphere) target, 
using the same instrumental settings as for the investigated sample. 
This additional step was performed either immediately before 
the sample cube acquisition, or after. The radiometric 
calibration then consists of a dark subtraction followed by the division of 
each sample monochrome image by the corresponding one of the Spectralon 
target.\\
In this study, both sample and calibration measurements were repeated 
4 times to ensure the stability and the quality of the calibrated data 
acquired across the wavelength domain investigated.
\subsection{Physikalisches Institute Radiometric Experiment 2 (PHIRE2)}
The PHIRE2 radio-goniometer was used to measure the phase curves for each 
of the prepared samples. As this instrument was presented in details in 
\citet{Pommerol_2011} and \citet{Jost_2017a}, only its main characteristics 
will be summarized hereafter.\\
The PHIRE2 setup uses an halogen lamp identical to the one of the MoHIS 
instrument. The light beam passes through a filter wheel, before being 
focused on an optical fibre. This filter wheels bears six bandpass filters 
centred at 450, 550, 650, 750, 905 and 1064 nm, with a bandpass 
width of 70 nm for the first four filters and of 25 nm for the last two 
NIR filters.\\
The optical fibre carries the filtered light beam to the illuminator arm 
of the goniometer, at the end of which a collimating lens and an iris 
combination allows to tune the beam's diameter between 5 and 20 mm onto 
the surface of the sample. A diameter of 5 mm was used for the 
measurements presented in this study. The setup allowed us to 
measure the sample' scattering properties for incidence and emergence 
angles ranging from -80\odeg to 80\odeg, and from azimuth angles varying 
from 0\odeg to 180\odeg.\\
The light scattered by the sample is measured by a silicon photodiode 
sensor, placed at the end of the goniometer's detector arm. Using 
complementary measurements of a 99\% reflective 
Spectralon$^{\text{\textregistered}}$ target, the recorded voltage is 
ultimately converted to a reflectance factor (REFF) value, similarly 
to what is done with MoHIS. We refer the reader to \citet{Pommerol_2011} 
and references therein for additional details on the radiometric 
calibration procedure.
\subsection{Photometric model}
\label{sec:phot_mdl}
For each sample investigated, we retrieve the best-fitting 
photometric parameters according to our modified implementation of the 
semi-empirical bidirectional reflectance model detailed in \citet{Hapke_1993}. 
Although based on the mathematically rigorous description of the radiative 
transfer occurring in a compact particulate medium composed of irregularly 
shaped and randomly oriented scatterers, the model incorporates empirical 
corrections, drawing from both theoretical and experimental considerations, 
to account for the superficial porosity of the medium, the unevenness of its 
surface and the non-linear reflectance surge near opposition (see 
\citealt{Feller_2016} and references therein, for more details). As in these 
previous studies, since the implementation of the model used here also draws 
on the remarks of \citet{Helfenstein_2011} and \citet{Shkuratov_2012}, 
we refer to it as the “HHS” model.\\

In the present study, in order to compare the results with the literature 
of cometary nuclei, we focused on modeling the obtained phase curves 
acquired at 550 nm, under 0\odeg of incidence and in terms of bidirectional 
reflectance (also referred to BRF below, and linked to the 
REFF by the relation BRF = cos(i)*REFF/$\pi$). As these measurements cover 
a phase angle range between 5\odeg and 75\odeg\hspace{-6pt}, the non-linear 
reflectance surge near opposition was modeled purely through the so-called 
shadow-hiding function \citep{Hapke_1986}.\\
Hence the measured reflectance values were modeled through the following 
function:
\begin{align*}
 &r_{HHS}(i,e,\alpha) = \frac{K\omega_{ssa}}{4\pi}\cdot\left(
 \frac{\mu_{0,e}(\bar{\theta})}{\mu_{0,e}(\bar{\theta})+\mu_{e}(\bar{\theta})}
                     \cdot S(i,e,\alpha, \bar{\theta})\right)\cdot \nonumber\\
 &\left[P_{HG}(\alpha,\omega_{ssa},\zeta)\cdot B_{SH}(\alpha,B_{SH,0}, h_{SH})+
 M_{scat}\left(\frac{\mu_{0}}{K},\frac{\mu}{K},\omega_{ssa}\right)\right]
     \stepcounter{equation}\tag{\theequation}\label{eq: HHS model}\\
\end{align*}
The variables in equation \ref{eq: HHS model} listed as i, e, 
and $\alpha$ are respectively the incidence, emergence and phase angles 
associated with a surface element of the sample and measured from the 
normal to the surface. $\mu_{0}$ and $\mu$ stand for the respective cosine 
of the incidence and the emergence angles.\\

The parameters of this photometric model are $\omega_{\text{ssa}}$, 
$\zeta$, B$_{SH}$, h$_{SH}$, $\bar{\theta}$ and K. They describe 
respectively the single-scattering albedo of the individual particles 
composing the medium, their scattering asymmetry factor, the amplitude and 
width of the shadow-hiding opposition effect (hereafter noted SH), the 
mean slope angle associated with the photometric roughness and finally the 
porosity coefficient. The description of these parameters and their bounds 
are summarized in Table \ref{tab:hapke-description}.\\

These variables and parameters are used in the single particle 
phase function (P$_{HG}$, modeled here by the single lobe Henyey-Greenstein 
function), the shadow-hiding function (B$_{SH}$), the multiple scattering 
contribution (M$_{scat}$), modeled here with the second order approximation 
of the Ambartsumian-Chandrasekhar H function detailed in 
\citealp{Hapke_2002}) and finally the shadowing function (S).
This last function together with the $\mu_{0,e}/(\mu_{0,e}+\mu_{e})$ 
term, replaces here the canonical Lommel-Seeliger disk function 
\cite{Seeliger_1887} and describes the scattering of light under a given 
viewing geometry by an uneven surface with a mean slope angle 
$\bar{\theta}$ \citep{Hapke_1984}.
Finally, following remarks expressed in \cite{Shkuratov_2012}, the 
exact analytical expression of the shadow-hiding function as given in 
\cite{Hapke_1993}, was used in our computations.\\

In this model, while the first four parameters can vary freely, 
the porosity coefficient (K) is bound to h$_{SH}$ through the empirical 
formula established from laboratory experiments by \cite{Helfenstein_2011} 
and given in Table \ref{tab:hapke-description}. Its associated domain of 
definition is therefore delimited by that of h$_{SH}$, the width of the 
shadow-hiding.\\
Additionally, we let the amplitude B$_{SH,0}$ vary freely between 
0 and 3. This upper bound was chosen as \cite{Hapke_1993} notes that should 
a medium be composed of clumps of small particles casting shadows on each 
other or particles with complex sub-particle sized structure casting 
shadows on itself, then one should expect the amplitude of the shadow-hiding 
to reflect this and be best-modeled with a value greater than unity. In our 
case, while exploring the best-fitting volume of parameters, we found this 
upper bound to close that parameter' space with enough margin for each and 
every sample.
\begin{table*}
\centering%
\footnotesize
\begin{tabular}{ccc}
{ Model parameter } & { Description } & { Bounds }\\
\hline 
{ $w_{SSA}$  }  & {Single-scattering albedo of the medium.} & {$\left] 0.0,1.0\right[ $}\\
\hline 
{ $\zeta$    }  & \makecell{Coefficient of the single lobe Henyey-Greenstein function \\
                             defined as the average cosine of emergence angle of      \\
                             the single particle phase function (SPPF). }
                 & { $\left]-1.0,1.0\right[ $}\\
\hline 
{$B_{SH, 0}$ }  & \makecell{Amplitude of the SH. $B_{SH,0}$ is a function of the Fresnel \\ 
                   reflection coefficient at normal incidence ($S(0)$).} {\par} 
                & { $\left[ 0.0,3.0\right] $}\\
\hline
{ $h_{SH}$  }   & \makecell{ Angular width of the SH. $h_{SH}$ is function of the medium's \\
                             porosity and its propensity to absorb and scatter light. }{\par}
                & { $\left[ 0.0,0.15\right] $}\\
\hline 
{ $\bar{\theta}$ } 
                & \makecell{The photometric roughness is the average macroscopic roughness \\
                            slope below the spatial resolution of the detector.} {\par}
                & { $\left[ 1^{\circ},60^{\circ}\right[ $}\\
\hline
{ $K$ }         & \makecell{Porosity coefficient, an empirical correction factor introduced    \\ 
                            to account for the role of superficial porosity in the scattering. \\
                            Our porosity coefficient is bound to $h_{SH}$ by the following     \\
                            empirical formula established by \cite{Helfenstein_2011}:          \\
    $K=1.069+2.109\cdot\text{h}_{SH}+0.577\cdot\text{h}_{SH}^{2}-0.062\cdot\text{h}_{SH}^{3}$ }
                & { $\left[ 1.07, 1.4\right] $}\\
\hline 
\end{tabular}
\caption{\label{tab:hapke-description} List and remarks on the model parameters used in this study}
\end{table*}
Further details and discussion on the HHS model can be found in 
\citet{Feller_2016} and \citet{Hasselmann_2016}.\\
For each measured phase curve, the fitting was performed through an 
iterative grid search with 21 steps per parameter, using the normalised 
root-mean-square deviation (RMS) as the estimator \citep{Li_2013} and 
evaluating its derivatives at each grid node to identify the best-fitting 
parameter sets.\\
The solutions presented below (Table \ref{tab:hpk_best_sols}) correspond 
to the average values from the 500 best-fitting solutions in the sense of 
the RMS. This limit was chosen to sift out the parameters volume containing 
the best solutions while retaining the same statistics for each dataset.\\

Finally, in order to allow further comparison with the literature, two 
integrated quantities, the geometrical and the bidirectional albedos, 
were computed. The geometrical albedo (A$_{geo}$) is defined as the ratio 
between the integrated bidirectional reflectance scattered towards the 
emerging hemisphere at 0\odeg of phase angle, and the corresponding 
brightness scattered by a Lambertian surface.\\
The bidirectional albedo (A$_{bd}$) is defined, for a given wavelength, 
as the integrated bidirectional reflectance scattered across the upper 
hemisphere given a uniform illumination from the incident hemisphere. The 
canonical Bond albedo is derived from the bidirectional albedo through the 
integration of that scattered brightness over all wavelengths.\\
Hence, these two quantities are computed as follows:
\begin{align}
    A_{geo} &= \pi\cdot \int_0^{\pi/2} r(e,e,0)\cdot \sin(2e)~\mathrm{d}e \\
    A_{bd}  &= \pi\cdot \iint_0^{\pi/2} r(i,e,\alpha)\cdot \sin(i)\cdot \sin(2e)~\mathrm{d}i\mathrm{d}e
    \label{eq:integrated_quantities}
\end{align}
Here, these two quantities were calculated based on the best-fitting 
solutions of the HHS model.
\FloatBarrier
\section{Spectro-photometric characterization}
\subsection{Spectral properties of the samples}
\label{sec:Spectroscopic measurements}
We present here the results derived from the spectro-imager's measurements of 
both the pure SiO$_{2}$ and juniper Charcoal samples as well as the prepared 
CoPhyLab dust mixtures. Their measured spectra are shown in Fig. 
\ref{fig:spectra_mix}, in terms of reflectance factor (REFF) and of reflectance 
normalised at 535 nm. This wavelength is close to the central wavelength 
of the standard Bessell V filter \cite{Bessell_1990} and it also corresponds 
to the central wavelength of the OSIRIS imager's NAC F23 filter of the ROSETTA 
mission, and will be discussed in the latter sections.
\subsubsection{Properties of the mixtures' end-members}
\begin{figure*}
  \begin{minipage}[h]{0.47\textwidth}
  \begin{center}
  \includegraphics[width=\linewidth]{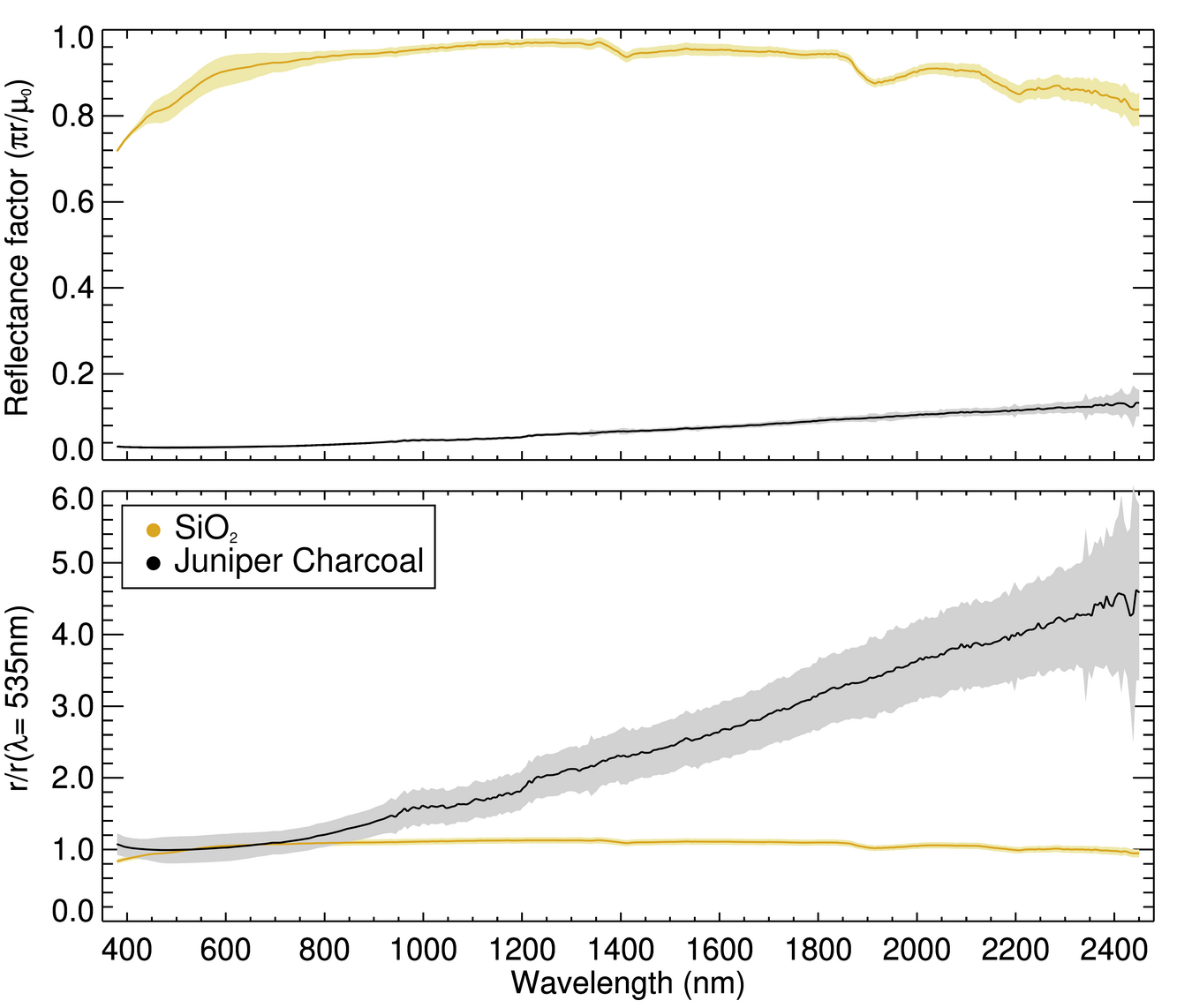}
  \end{center}
  \end{minipage}
  \hfill
  \begin{minipage}[h]{0.47\textwidth}
  \begin{center}
  \includegraphics[width=\linewidth]{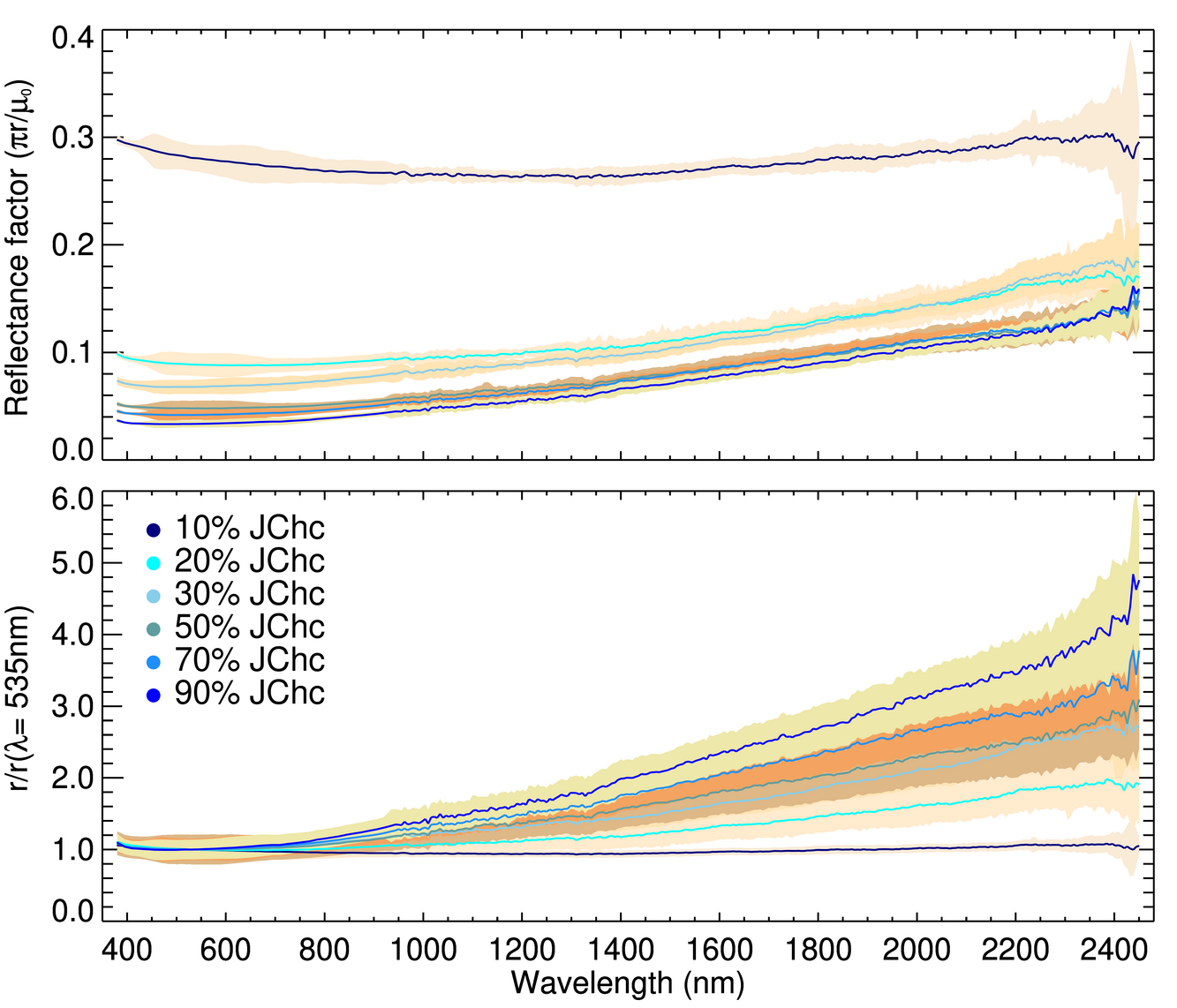}
  \end{center}
  \end{minipage}
  \caption{\label{fig:spectra_mix} Spectra of both end-members and 
     mixtures measured with MoHIS are plotted above. Each spectrum was 
     acquired near opposition ($\alpha$ $\sim$ 4 to 7\odeg)
     For each spectrum, the propagated dispersion associated to the 
     measurements is depicted by a corresponding filled color envelope 
     around the solid lines. For clarity, only the spectra for the 10\%, 
     20\%, 30\%, 50\%, 70\% and 90\% juniper charcoal mixtures are displayed 
     in the plots of the right column. --
     Top plots: Reflectance spectra for the SiO$_2$ and JChc samples (left) 
     and for Cophylab dust mixtures (right). -- 
     Bottom plots: Reflectance spectra normalised at 535 nm for the SiO$_2$ 
     and JChc samples (left) and for Cophylab dust mixtures (right).} 
\end{figure*}
Fig. \ref{fig:spectra_mix}, top-left) shows the extreme contrast between 
the spectra of SiO$_2$ and juniper charcoal. The juniper charcoal spectrum 
presents low REFF values from the visible to the near-infrared (from 
0.031\mmp 0.003 at 380 nm through 0.047\mmp 0.007 at 1000 nm to 0.13\mmp 
0.06 at 2450 nm), an absence of marked spectral features and a nearly 
monotonic behaviour. Although the charcoal REFF spectrum marginally 
decreases to 0.029\mmp 0.007 at 535 nm, it increases readily and steadily 
from 600 nm onward and across the whole near-infrared domain. These 
spectral features are consistent with similar carbon-bearing materials 
\citep{Cloutis_1994, Moroz_1998, Andres_2005}.\\
Using the spectral slope formula from \cite{Fornasier_2015}, and adapting 
for the nearest-neighbours wavelengths, we find that this monotonic 
behaviour is marked with a stronger spectral slope between 1000 nm and 
2400 nm (S $\sim$ 13\mmp 4\%/100 nm) than between 380 nm and 1000 nm 
(S $\sim$ 8\mmp 2 \%/100 nm). This behaviour is particularly obvious when 
the spectrum is normalised at 535 nm (see Fig. \ref{fig:spectra_mix}, 
lower-left), with the reflectance values in the near-infrared range 
increasing steadily, to reach over four times the reflectance at 535 nm 
(REFF$_{535 mn}$ $\simeq$ 0.029\mmp 0.007) at wavelengths longer than 
2200 nm.\\
Moreover, the normalised spectrum highlights the presence of a wide but 
shallow dip in reflectance ($\sim$ 5\% with respect to the surrounding 
continuum) extending from $\sim$ 1000 nm to $\sim$ 1200 nm. This wavelength 
range matches the wavelength windows for overtones of the CH$_{x}$ 
absorption bands \citep{Lipp_1992}, such as those exhibited by polycyclic 
aromatic hydrocarbons (e.g. \citealt{Izawa_2014}).\\

On the contrary, the SiO$_2$ spectrum exhibits much higher reflectance 
factor values (see Fig. \ref{fig:spectra_mix}, top-left), and displays 
several spectral features. The SiO$_2$ spectrum increases steadily 
throughout the visible domain (from a REFF value of 0.73\mmp 0.01 at 
380 nm to 0.96\mmp 0.02 at 800 nm) and reaches a maximum value of 
0.99\mmp 0.02 near 1230 nm, before gradually decreasing to a value of 
0.83\mmp 0.07 at 2450 nm.\\

We interpret this reflectance decrease across the infrared domain 
as a likely effect of the grain-size distribution of this weekly absorbent
particulate medium. Indeed, numerous particles present a diameter 
smaller than 1$\mu$m in the SEM images, (see Fig. \ref{fig:sem_sio2}). At 
these wavelengths and for such grain-size, individual particles are expected 
to diffract light and behave as strong absorbers. However, given the 
compacity of this SiO$_2$ particulate medium, the clear presence of 
agglomerates in the SEM images and the SiO$_{2}$'s negligible absorption 
index in this wavelength range \citep{Kitamura_2007}, it is reasonable 
to assume that part of the medium is in a volume scattering regime in which 
diffraction is negligible, as clumps of small-than-wavelength 
particles scatter light in the way a single particle with an equivalent 
diameter would \citep{Hapke_1993, Mustard_1997, Rousseau_2017}.\\

Furthermore, in the infrared domain, the SiO$_2$ spectrum displays
three dips in reflectance: at $\sim$ 1400 nm ($\sim$ 7143 cm$^{-1}$), at 
$\sim$ 1900 nm ($\sim$ 5263 cm$^{-1}$), and at $\sim$ 2200 nm ($\sim$ 4545 
cm$^{-1}$). These spectral bands are associated with water and/or hydroxyl 
group from weathering of the sample and/or adsorption of atmospheric water 
vapor (\citealt{Elliott_1971, Lipp_1992, Workman_2012}, and references 
therein). Additional spectral measurements were performed after 
placing the SiO$_{2}$ sample in a desiccating oven at 100\odeg C at room 
pressure for 30 hours, in which the spectral features were still present. 
We note, according to \citealt{Zhuravlev2000} and \citealt{Rahman_2008}, 
that the total removal of -OH and H$_2$O bound groups from amorphous 
SiO$_2$ start to occur from temperatures higher than 150\odeg C at room 
pressure.\\
Although investigating the water vapor adsorption efficiency of 
this SiO$_2$ sample would have been beyond the scope of this analysis, 
we interpret the continuous presence of these hydration and hydroxylation 
bands as a likely indication of -OH and H$_2$O strongly tied to the 
SiO$_2$ matrix, both of which might also contribute to the predisposition 
of the SiO$_{2}$ sample to form aggregates.\\

While some coals and charcoals (e.g. \cite{Moroz_1998}, 
\citet{Pommerol_2008}) display hydration and hydroxylation features 
in their spectra, our JChc sample does not exhibit such spectral features. 
Although \citet{Dias_2016} reports that the water content of a charcoal 
sample is related to its porosity, \citet{Quirico_2016} also points out 
that the absence of hydration or hydroxylation features in the 
near-infrared is a sign of charcoal maturity, and a low hydrogen to carbon 
atoms (H/C) ratio.\\
Both the studies of \citet{Moroz_1998} and \citet{Quirico_2016} highlight,  
respectively through the examples of the anthraxolites and the anthracite 
sample PSOC-1468, how the longer a coal goes through a carbonisation 
process, the greater its carbon content and the lower its overall 
reflectance become.\\
Together with the broad reflectance dip across the 1000 -- 1300 nm 
range, we therefore interpreted the absence of distinguishable hydration 
and hydroxylation features in the juniper charcoal spectrum as a likely 
indication of an elevated carbon-content and a low H/C ratio.\\

In order to verify this interpretation, elemental analyses of juniper 
charcoal samples were performed using a Thermo Scientific FLASH 2000 
organic elemental analyser (Brechbühler AG, Schlieren, Switzerland). 
This instrument, equipped with a reactor consisting of chromium 
oxide, elemental copper, and silvered cobaltous-cobaltic oxide, allowed 
to measure the amount of carbon, hydrogen and nitrogen present in the 
provided samples. Atropine, cystine, and sulphanilamide, provided by 
Elemental Microanalysis (Okehampton, UK), were used as standards for 
calibration, while sample quantities in the range of 1-3 mg were weighed 
for analysis.\\
Two different samples of juniper charcoal investigated with this instrument 
contained 80.42\mmp 0.43\% of carbon and 2.69\mmp 0.01\% of hydrogen, and 
thus a H/C ratio of $\sim$ 3\%. No trace of nitrogen was found. Such a 
value of H/C is comparable to those of anthraxolites-rich coals, which 
also present a featureless monotonically increasing spectrum and are 
generally associated with polyaromatic compounds (\citealt{Moroz_1998} 
and references therein).\\

\subsubsection{Intimate mixtures of SiO$_{2}$ and juniper charcoal}
Similarly, the REFF spectra of the intimate mixtures without and with 
normalisation at 535 nm have been plotted in Fig. \ref{fig:spectra_mix} 
(top- and bottom-right). For better readability, only a selection of the 
measured spectra is plotted in that figure.\\
The direct comparison of the pure SiO$_{2}$ REFF spectrum (Fig. 
\ref{fig:spectra_mix}, top-left) with that of the 90\% SiO$_{2}$/10\% 
JChc mixture (Fig. \ref{fig:spectra_mix}, top-right) clearly shows how 
the inclusion of a limited amount (by mass) of juniper charcoal 
drastically affects the reflectance of the mixture: whereas the REFF 
values of the SiO$_{2}$ spectrum were above 0.75 across the considered 
wavelength range, the 10\% juniper charcoal mixture presents REFF values 
between 0.25 and 0.30 across the same range (i.e. in average a 67\mmp 4\% 
reduction).\\
However, while compared to the pure SiO$_{2}$ spectrum, each increment of 
juniper charcoal lowers the average REFF value further, the average 
difference between the spectra of the previous and the new increment 
decreases. The specific progression at 535 nm, listed in Table 
\ref{tbl:reff_535}, can be fitted with an exponential reduction, and is 
investigated in the discussion section.\\
\begin{table}
  \centering
  \caption{\label{tbl:reff_535} Variation of the mixtures REFF values and the 
           pure juniper charcoal value with respect to the pure SiO$_{2}$ REFF 
           value at 535 nm ($\sim$ 88.5\mmp 0.6\%).}
  \begin{tabular}{|c|c||c|c|}
    \hline
    Sample & REFF ratio & Sample & REFF ratio \\
    \hline
    10\% JChc & 70\mmp 4\% &  60\% JChc & 95\mmp 1\%     \\
    20\% JChc & 90\mmp 2\% &  70\% JChc & 96\mmp 1\%     \\
    30\% JChc & 93\mmp 2\% &  80\% JChc & 96\mmp 1\%     \\
    40\% JChc & 94\mmp 2\% &  90\% JChc & 96\mmp 1\% \\
    50\% JChc & 95\mmp 2\% & 100\% JChc & 97\mmp 1\%     \\
    \hline
  \end{tabular}
\end{table}
Additionally, these spectra do not exhibit the hydration and hydroxylation 
spectral features present in the SiO$_{2}$ spectrum, even for the mixture 
containing only 10 wt. \% of juniper charcoal. This observation is consistent 
with the literature (e.g. \citealt{Pommerol_2008, Jost_2017, Rousseau_2017}). 
The authors of these studies found that the reflectance properties of a 
binary mixture composed of both very bright and very dark materials would 
be dominated by the dark end-member.\\

The comparison of these spectra with that of juniper charcoal points out 
that, while mixtures with a charcoal content higher than 10\% exhibit a 
profile akin to that of the pure juniper charcoal sample, the spectrum 
of the 10\% charcoal mixture differs from those of both pure --
end-members. The 10\% juniper charcoal spectrum follows a shallow u-curve 
profile, with a marked negative spectral slope between 380 nm and 800 nm 
(-2.3\mmp 0.2 \%/100 nm) and a small positive spectral slope  between 1200 
nm and 2400 nm (1.0\mmp 0.2 \%/100 nm).\\
The spectra of samples with a larger charcoal content match the trend of 
the pure juniper charcoal sample, with REFF values slightly decreasing 
between 400 and 600 nm and then increasing monotonously up to 2450 nm. 
Furthermore, across the 380 nm -- 800 nm range, we note that the mixtures' 
spectral slope increases in pair with the juniper charcoal content up to 
a value of 1.2\mmp 0.4 \% /100 nm for the mixture with a JChc mass 
fraction of 90\%. Similarly, the spectral slopes across the 1200 mm -- 
2400 nm also increase steadily up to 12\mmp 6\% / 100 nm for the 90\% 
charcoal mixture, with a trend of 0.21\mmp 0.02 \%/100 nm/\% of charcoal 
content.\\

We interpret the absence of the hydration and hydroxylation bands from 
the mixtures' spectra, the rapid overall decrease in reflectance, as well 
as the progressive matching of spectral slopes across both the visible 
and the near-infrared range as an further indication of the juniper 
charcoal fraction controlling the spectroscopic and photometric, and thus 
more generally the optical properties of the mixtures, as discussed in 
\citet{Rousseau_2017}. We therefore interpret the features distinguishing 
the 10\% juniper charcoal spectrum from the other mixtures' spectra (e.g. 
the high REFF values and the low spectral slopes values) as the expression 
of the transition from the SiO$_2$ fraction driving the optical properties 
of the mixture to them being controlled by the juniper charcoal fraction.\\
\FloatBarrier
\subsection{Phase curves and photometric modeling}
\label{sec:appendixmodel}

\begin{figure*}
\begin{minipage}[h]{0.39\textwidth}
  \begin{center}
    \includegraphics[width=\linewidth]{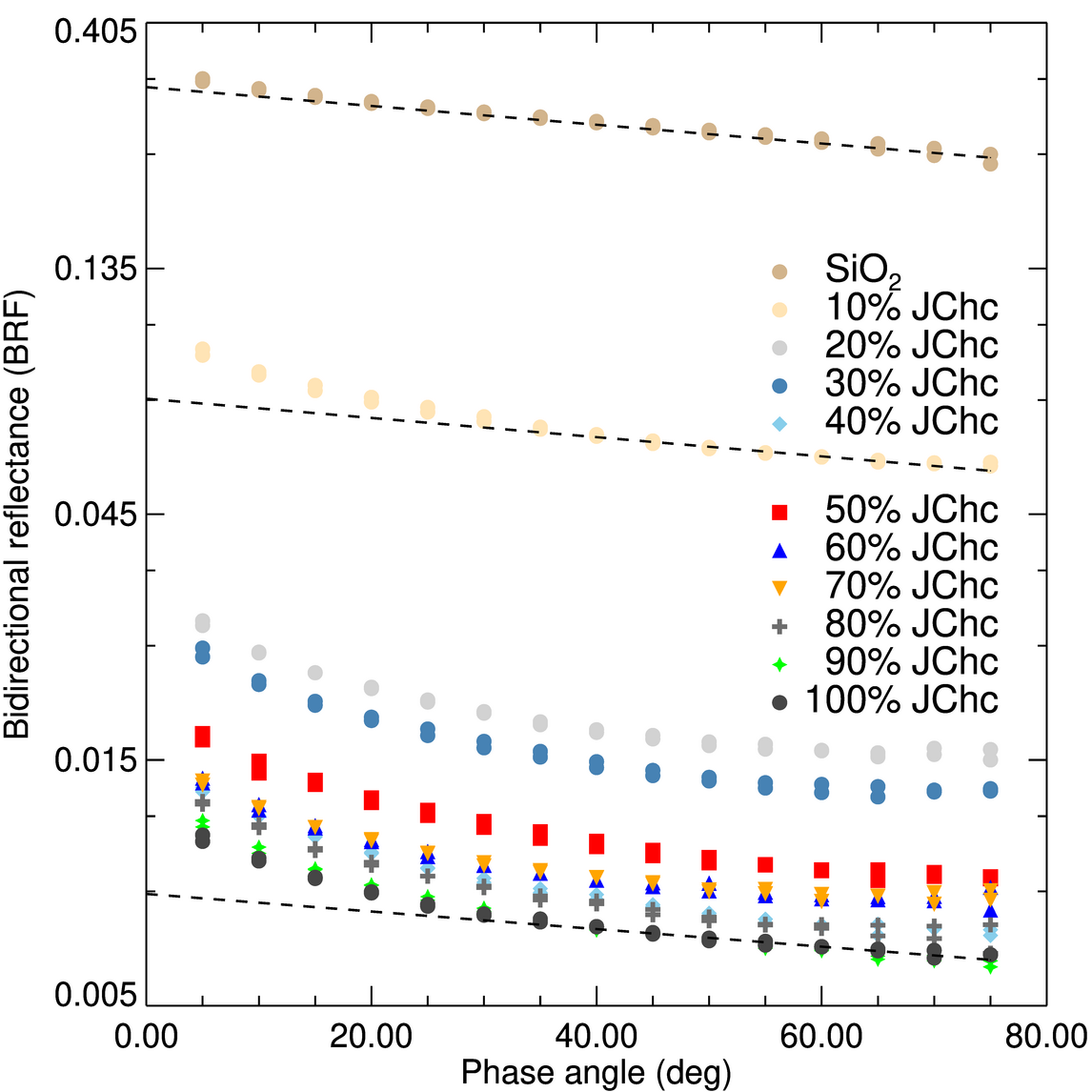}
  \end{center}
\end{minipage}
\hfill
\begin{minipage}[h]{0.58\textwidth}
  \begin{center}
\includegraphics[width=\linewidth]{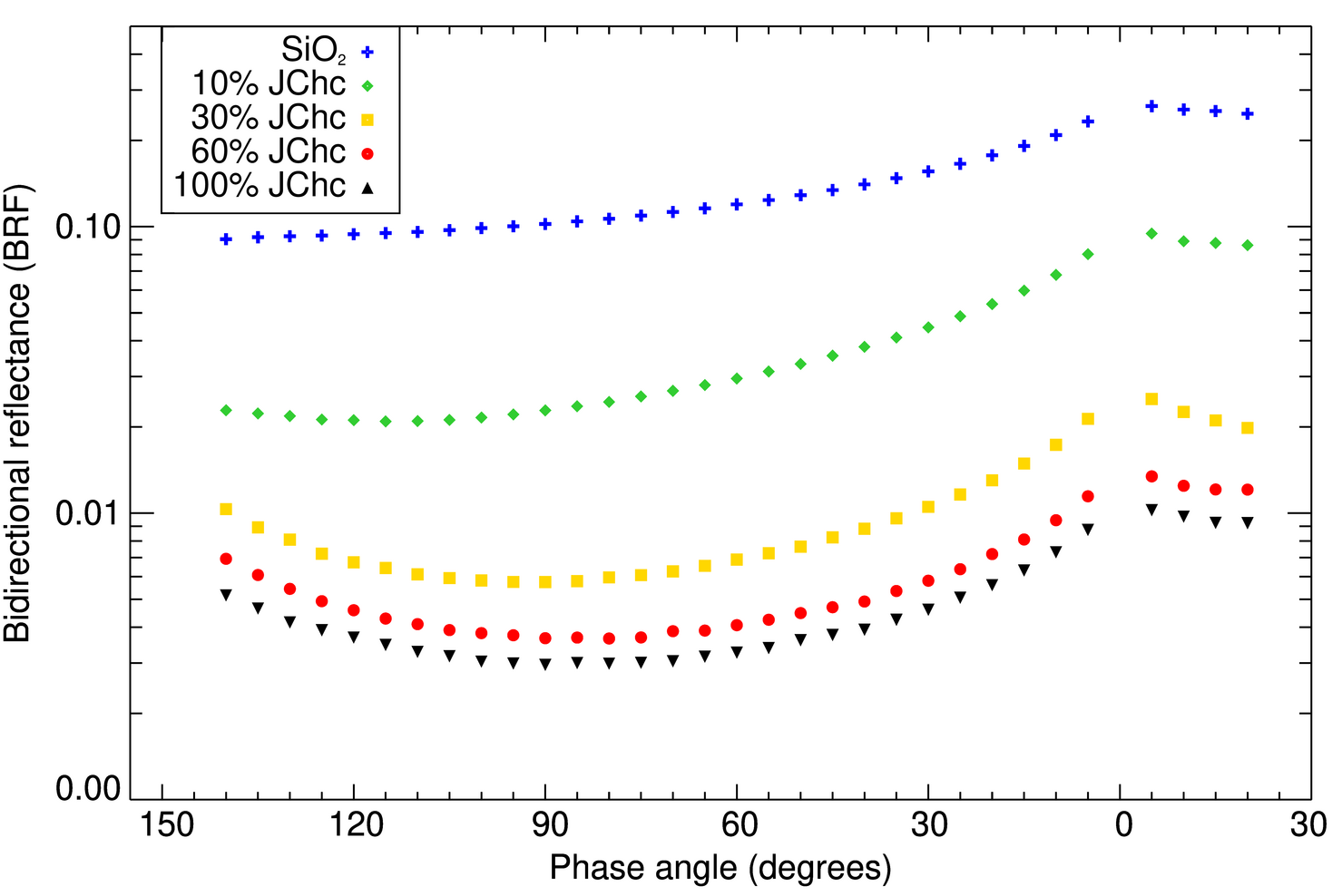}
  \end{center}
\end{minipage}
\caption{\label{fig:phot_phase_curves} Phase curves of the pure 
   materials and their mixtures at 550 nm, at an incidence angle of 
   0\odeg (left) and 60\odeg (right). For clarity, the dispersion of 
   the measurements are not included, and the reflectance axis follows 
   a logarithmic scale. -- 
   Left: The black dotted lines plotted over the phase curves of the 
   end members and the 10\% JChc mixture are the exponential fits of 
   the 30\odeg-75\odeg portion of the corresponding curves, and 
   highlight the non-linear surge of the reflectance at low phase 
   angles.}
\end{figure*}

Selections of measured phase curves at 0\odeg and 60\odeg of 
incidence are plotted in Fig \ref{fig:phot_phase_curves}a and 
\ref{fig:phot_phase_curves}b, respectively.
Overall, the phase curves in Fig. \ref{fig:phot_phase_curves} 
show a strong decrease in reflectance with the increase of charcoal 
content. They also display a non-linear reflectance surge at low 
phase angles, typically associated to self-shadowing in a particulate 
medium, also known as the shadow-hiding opposition effect (SHOE, 
\citealt{Hapke_1993}). In Fig. \ref{fig:phot_phase_curves} (left), we 
note a difference in the magnitude of the departure from linear 
behaviour between the pure SiO$_2$ sample and the darker samples, 
suggesting a difference in the amplitudes of the SHOE affecting each 
sample.\\
As discussed previously, we fitted the acquired phase curves at 550 
nm and under an incidence angle of 0\odeg with the implementation of 
the Hapke photometric model discussed in section \ref{sec:phot_mdl}. 
Focusing on this part of the dataset has a drawback, as the constraints 
on the asymmetry factor ($\xi$) and on the photometric roughness 
($\bar{\theta}$) are then limited (see e.g. \citealt{Pilorget_2013} 
and \citealt{Labarre_2017}). However, to allow direct comparison with 
the literature (e.g. \citealt{Li_2013}, \citealt{Fornasier_2015}), we 
ultimately focused on these particular phase curves.\\

The sets of best-fitting parameters for each sample are listed in 
Table \ref{tab:hpk_best_sols} and the associated quality-fits are 
presented in fig. \ref{fig:full_phase_curves}. We find that while all 
materials and samples present a back-scattering behaviour, the SiO$_2$ 
sample is more back-scattering than the juniper charcoal sample, with 
its corresponding asymmetry parameter ($\xi_{SiO_2}\simeq$ -0.28) being 
almost twice as large as that of the juniper charcoal ($\xi_{JChc}\simeq$ 
-0.15). Compared to the literature and assuming no major dependencies of 
this parameter with wavelength, we find that these samples are slightly 
less back-scattering than most cometary nuclei, for which the 
best-fitting values are currently found to be between -0.5 and -0.3 
(\citealt{Ciarniello_2015, Feller_2016, Hasselmann_2017} and references 
therein).\\

The contrast between the SiO$_2$ and the juniper charcoal also 
appears through the derived parameters of the SHOE function (B$_{SH,0}$ 
and h$_{SH}$) which point to two different photometric behaviours near 
opposition. Both the SiO$_2$ sample and the 10\% charcoal mixture 
present a moderate reflectance surge (B$_{SH,0} <$ 1) with peaks having 
an half-width at half-maximum (HWHM $\simeq2\cdot$h$_{SH}$, 
\citealp{Hapke_1993}) smaller than 10\odeg\hspace{-6pt}, while most mixtures 
and the pure juniper charcoal sample each exhibit a reflectance surge with 
a wider peak (HWHM $\simeq$ 10\odeg -- 17\odeg\hspace{-6pt}), and a large 
amplitude (B$_{SH,0} >$ 1). These two behaviours and the dominant 
role of the juniper charcoal are consistent with our previous observations 
of the SiO$_2$ material being bright and weakly absorbing (thus casting 
small shadows), while the juniper charcoal is dark and strongly absorbing 
and thus cast stronger shadows). These results are also consistent with the 
literature for opaque surfaces (e.g. \citealt{Shepard_2007, 
Shevchenko_2012, Masoumzadeh_2016}), on which the SHOE function is 
the main driver behind their reflectance surge at opposition.\\ 

Additionally, the larger-than-one SHOE amplitude are also 
consistent with the literature on atmosphereless small-bodies (e.g. 
\citealt{Helfenstein_1994, Simonelli_1998, Li_2004, Li_2013, 
Feller_2016}). We note indeed that in the canonical description of the 
Hapke photometric model, the B$_{SH,0}$ parameter stands for the 
amplification of the shadow-hiding phenomenon due to the effect of the 
size of the scattering particle. However, it is indicated in 
\cite{Hapke_1993} that particles with a complex geometry and their 
agglomerates are expected to cast shadows on themselves ("self shadowing"), 
which would be then manifest through values higher than 1. Hence, we 
interpret the higher-than-one B$_{SH,0}$ values to further point to the 
driving influence of the charcoal given the numerous presence of agglomerates 
and opaque particles complex shapes as observed in the SEM images.\\

Considering \citet{Helfenstein_2011} in the implementation of this 
photometric model, the width of the SHOE function is tied with the porosity 
correction factor K and therefore to the superficial porosity of the medium, 
as described in \citet{Hapke_2008}. We thus derived  the superficial porosity 
values (P) listed in Table \ref{tab:hpk_best_sols} from the best-fitting 
h$_{SH}$ values. While slightly higher than the bulk porosities measured 
with the helium pycnometer (P$_{SiO_2}\simeq$ 65\%, P$_{JChc}\simeq$ 68\%), 
these values theoretically reflect the porosities in the uppermost 
layers of scatterers. They are consistent with the other measures of this 
study, with measurements of 67P/CG derived from Rosetta's optical 
observations \citep{Fornasier_2015, Hasselmann_2017} or from Philae's radar 
\citep{Kofman_2015}, and with laboratory experiments on aggregates 
\citep{Bertini_2007, Lasue_2011}.\\

The photometric roughness describes the average slope of the sample surface 
in the model, as perceived below the spatial resolution of the detector (at 
the centimetre scale given the setup of these measurements). The best-fit 
values are $\bar{\theta} \simeq $ 4\odeg-11\odeg. These values are slightly 
lower than the retrieved photometric roughnesses of visited cometary nuclei 
observed at larger spatial resolutions (e.g. $\bar{\theta}_{81P/Wild-2} 
\simeq$ 15\odeg, \citealt{Li_2013}), and than the global and local 
roughnesses found for 67P/C-G (between 15\odeg and 33\odeg, see 
\citealt{Fornasier_2015, Feller_2016, Hasselmann_2017}). However, these 
values as such are consistent with loose and smooth-surfaced particulate 
media with grain sizes between 0.1 and 50 $\mu$m investigated in laboratory 
settings \citep{Shepard_2007}.
\addtolength{\tabcolsep}{-3pt}
\begin{table*}
 \centering
  \caption{\label{tab:hpk_best_sols} List of best-fitting parameters for the HHS
         photometric model (averaged over the best 500 solutions) and associated
         dispersion. See Table \ref{tab:hapke-description} for a description of 
         each model parameter. The porosity (P) is derived from the porosity 
         coefficient K. 
         $\chi^{2}$ and R$^2$ standing respectively as the normalised
         root-mean-square deviation and the Pearson coefficient. A$_{geo}$ and 
         A$_{bd}$ respectfully corresponds to the geometrical and bidirectional
         albedos computed from the associated best-fitting parameters.}
  \begin{footnotesize}
  \begin{tabular}{|c|c|c|c|c|c|c|c|c||c|c|}
   \hline
    Samples & $\omega_{ssa}$ & $\xi$ & B$_{SH,0}$ & h$_{SH}$ & $\bar{\theta}$ (deg) & P (\%) & $\chi^{2}$ (x10$^{-3}$) & R$^2$ & A$_{geo}$ & A$_{bd}$\\
   \hline
   Pure SiO$_2$ & 0.975\mmp 0.1 & -0.276\mmp 0.003 & 0.59\mmp 0.02 & 0.07\mmp 0.01 & 5.\mmp 1. & 87.\mmp 1. & 2.070 & 0.994 & 1.07\mmp 0.01  & 83.2\mmp 0.9 \\
   10\% charcoal& 0.478\mmp 0.3 & -0.241\mmp 0.002 & 0.90\mmp 0.01 & 0.09\mmp 0.01 &  $<$ 5.*  & 83.\mmp 1. & 0.595 & 0.997 & 0.34\mmp 0.01  & 23.4\mmp 0.3 \\
   20\% charcoal& 0.139\mmp 0.2 & -0.178\mmp 0.003 & 1.61\mmp 0.03 & 0.11\mmp 0.01 & 6.\mmp 1. & 79.\mmp 1. & 0.176 & 0.997 & 0.11\mmp 0.01  &  6.4\mmp 0.1 \\
   30\% charcoal& 0.117\mmp 0.1 & -0.186\mmp 0.001 & 1.64\mmp 0.01 & 0.11\mmp 0.01 & 4.\mmp 1. & 79.\mmp 1. & 0.231 & 0.995 & 0.09\mmp 0.01  &  5.5\mmp 0.1 \\
   40\% charcoal&  0.062\mmp 0.2 & -0.190\mmp 0.006 & 1.57\mmp 0.09 & 0.13\mmp 0.01 & 6.\mmp 1. & 76.\mmp 1. & 0.083 & 0.998 & 0.05\mmp 0.01  &  2.9\mmp 0.2 \\
   50\% charcoal&  0.09\mmp 0.1 & -0.217\mmp 0.002 & 1.24\mmp 0.02 & 0.09\mmp 0.01 & 4.\mmp 1. & 82.\mmp 1. & 0.169 & 0.994 & 0.06\mmp 0.01  &  3.8\mmp 0.1 \\
   60\% charcoal&  0.076\mmp 0.2 & -0.165\mmp 0.006 & 1.44\mmp 0.05 & 0.11\mmp 0.01 & 7.\mmp 1. & 79.\mmp 1. & 0.141 & 0.993 & 0.05\mmp 0.01  &  3.3\mmp 0.1 \\
   70\% charcoal&  0.076\mmp 0.1 & -0.171\mmp 0.002 & 1.33\mmp 0.03 & 0.11\mmp 0.01 &  $<$ 5.*  & 78.\mmp 1. & 0.099 & 0.996 & 0.050\mmp 0.009 & 3.4\mmp 0.1 \\
   80\% charcoal&  0.063\mmp 0.2 & -0.180\mmp 0.008 & 1.46\mmp 0.08 & 0.12\mmp 0.01 &10.\mmp 1. & 77.\mmp 2. & 0.147 & 0.992 & 0.046\mmp 0.008 & 2.8\mmp 0.2 \\
   90\% charcoal&  0.057\mmp 0.2 & -0.179\mmp 0.007 & 1.49\mmp 0.08 & 0.12\mmp 0.01 & 9.\mmp 1. & 76.\mmp 1. & 0.076 & 0.997 & 0.042\mmp 0.005 & 2.5\mmp 0.2 \\
   Pure charcoal&  0.056\mmp 0.3 & -0.155\mmp 0.006 & 1.40\mmp 0.05 & 0.15\mmp 0.01 &11.\mmp 2. & 73.\mmp 2. & 0.070 & 0.997 & 0.038\mmp 0.002 & 2.5\mmp 0.2 \\
   \hline
   \hline
9P/Tempel 1$^{1}$& 0.039\mmp 0.005 & -0.49\mmp 0.02 & [1]            & [0.01]          & 16\mmp 8     & -   & - & - & 0.056 & 0.013 \\
    67P/C-G$^{2}$& 0.026\mmp 0.001 & -0.42\mmp 0.02 & 2.56\mmp 0.067 & 0.067\mmp 0.005 & 16.3\mmp 0.2 & 85. & - & - & 0.052 & - \\
   \hline
Average C-type$^{3}$& 0.025      & -0.47          & 1.03        & 0.025        & 20   & - & - & - & 0.049 & - \\
    Deimos$^{4}$ & 0.079\mmp 0.002 & -0.29\mmp0.03  & 1.6\mmp 0.6 & 0.07\mmp 0.3 & 16.4 & - & - & - & 0.067\mmp0.007 & 0.027\mmp0.003\\
    Phobos$^{4}$ & 0.054         & -0.13          & 5.7         & 0.072        & 22   & - & - & - & 0.071\mmp0.012 & 0.021\mmp0.004\\
   \hline
  \end{tabular}
    \newline
  \end{footnotesize}
  \begin{flushleft}
  \vspace*{-1em} *: Upper limit of the subset of optimal solutions.\\
  1: \citet{Li_2007}, parameters at 550 nm (with B$_{SH,0}$ and h$_{SH}$ fixed); 2: \citet{Feller_2016}, parameters at 535 nm;\\
  3: \citet{Helfenstein_1989}, parameters in the V band; 4:\citet{Thomas_1996, Simonelli_1996}, parameters at 540 nm.
  \end{flushleft}
\end{table*}
\addtolength{\tabcolsep}{3pt}
\section{Discussion}
\label{sec:discussion}
\subsection{Influence of the dark end-member}
\begin{figure*}
  \begin{minipage}[h]{0.47\textwidth}
  \begin{center}
  \includegraphics[width=\linewidth]{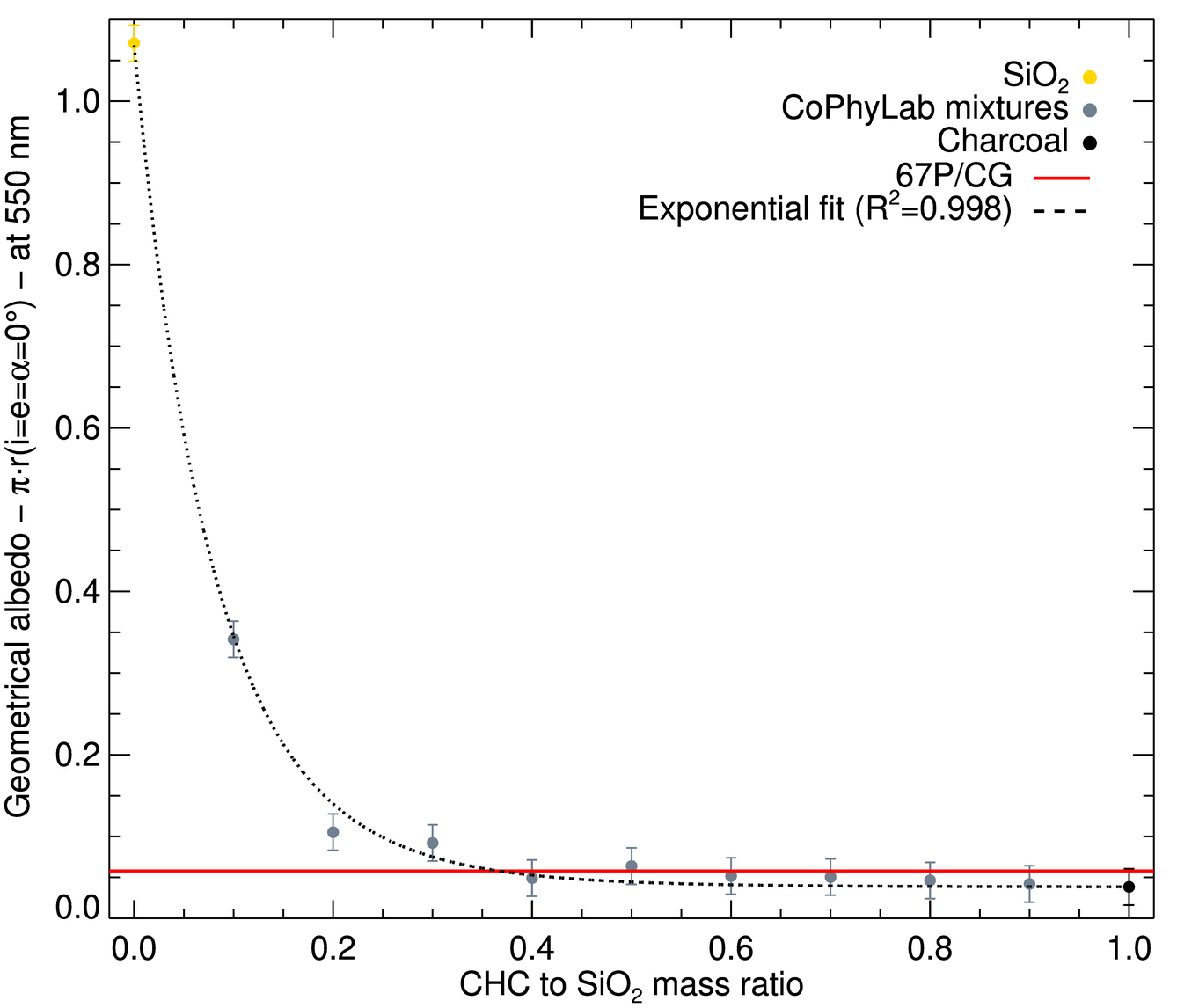}
  \end{center}
  \end{minipage}
  \hfill
  \begin{minipage}[h]{0.47\textwidth}
  \begin{center}
  \includegraphics[width=\linewidth]{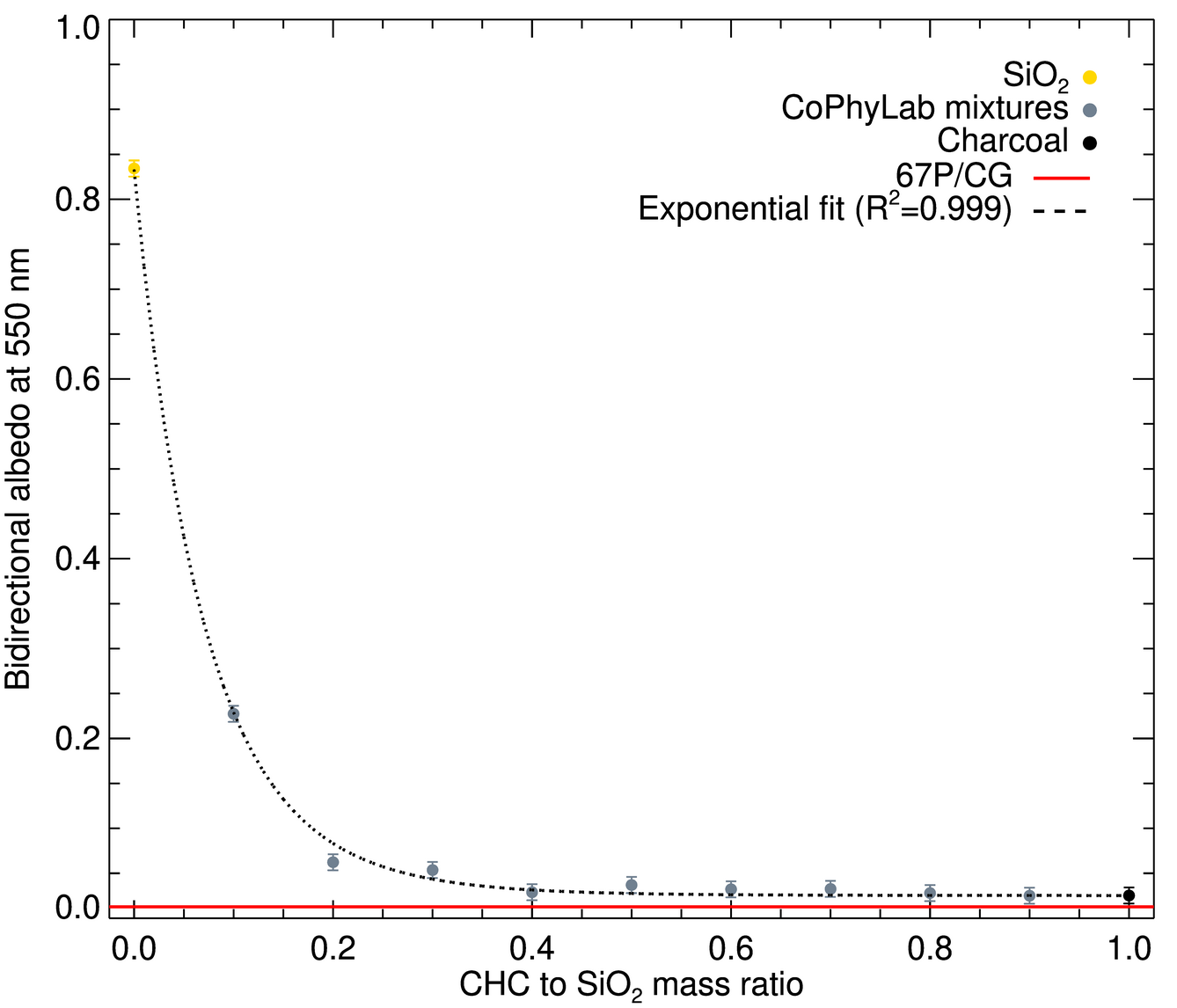}
  \end{center}
  \end{minipage}
  \caption{\label{fig:albedo_exp} Left: Geometric albedos values for the 
     CoPhyLab dust mixtures and its end-members. -- Right: Bidirectional 
     albedos values for the CoPhyLab dust mixtures and its end-members. -- 
     As illustrated in both plots, the strong albedo decreases are best 
     fitted with simple exponential functions (black dashed lines). The red line 
     stands for the corresponding albedo of 67P/Churyumov-Gerasimenko's 
     nucleus at 550 nm (Fornasier \textit{et al.}, 2015).}
\end{figure*}
As detailed throughout the Results section, both spectroscopic and photometric 
measurements point to the prominent role of the ground juniper charcoal in 
the resulting optical properties of the CoPhyLab dust mixtures. This 
observation is in agreement with the literature (e.g. \citealt{Warren_1980, 
Clark_1984, Pommerol_2008, Yoldi_2015}). In each of these studies, the 
authors found that for mixtures of bright and dark materials, the inclusion 
of a fraction of the dark end-member was sufficient to strongly reduce the 
overall brightness of the mixture. Using $\mu$m-sized particles, 
\citet{Cloutis_2015} reported specifically that such decrease is more 
intense for intimate mixtures than for areal mixtures. Furthermore, 
\citet{Rousseau_2017} has shown that this decrease was also observed when 
mixing sub-$\mu$m-sized particles. The originality here, further 
discussed in the rest of this section is that we are in principle in the 
opposite configuration where the particles of the dark material are much 
bigger in size than the individual particles of the bright material.\\

Both \citet{Jost_2017} and \citet{Rousseau_2017} discussed the mechanisms of 
such reflectance decreases in either the context of icy or non-icy mixtures. 
In the case of intimate mixtures of $\mu$m-sized water-ice and activated 
charcoal, \cite{Jost_2017} reported that the variations of the hemispherical 
albedos appeared to follow an exponential decrease as the amount of activated 
charcoal was increased.\\
We list in Table \ref{tab:hpk_best_sols} the bidirectional and 
geometric albedo values plotted in fig. \ref{fig:albedo_exp}. Similarly to 
these two studies, we also find that their progression can be 
satisfactorily fitted with an exponential function of the type albedo = 
albedo$_{\text{JChc}}$+$e^{\left(a+b\cdot (1+\omega_{\text{JChc}})\right)}$, 
with $\omega_{\text{JChc}}$ the mass fraction of the charcoal. For the 
geometric albedo values, the best set of fitting parameters (a, b) was found 
to be equal to (0.03 \mmp 0.01,-12.7 \mmp 0.1) and in the case of the 
bidirectional albedo to (-0.21 \mmp 0.01, -14.5 \mmp 0.1).\\
Considering the albedo values of Table \ref{tab:hpk_best_sols}, and 
given the best-fitting parameters, we find that all mixtures with at least 
40\% of charcoal by mass present a geometrical albedo close to or lower than 
that of 67P/C-G's nucleus (A$_{geo, 550nm}\simeq$ 5.9\mmp0.3\%).
\FloatBarrier
\subsection{Cross-sectional scattering efficiencies and surface 
heterogeneities}
Although the effective grain-size distribution of particles at the surface of 
comet 67P/C-G has not been characterized all across the nucleus, the physical 
properties of both ice and dust fractions have been investigated. 
Freshly exposed water-ice-enriched material was found to be best 
reproduced with micrometric to millimetric water-ice grains (e.g. 
\citealt{DeSanctis_2015, Filacchione_2016, Raponi_2016}). Ejected dust 
particles in the inner coma were investigated both remotely with the 
spectrometer VIRTIS and under the microscope with the instrument MIDAS.
\citet{Bockelee-Morvan_2017} and \citet{Mannel_2019} notably reported the 
observations of sub-micrometric to micrometric-sized dust grains and 
aggregates. \citet{Rousseau_2017} found, through laboratory experiments, a 
satisfactory match for 67P/C-G' spectral properties in the visible domain, by 
mixing sub-micrometric grains of mature coal, pyrrhotite, and dunite.\\
Further laboratory measurements presented in \citet{Jost_2017a} showed that
the sublimation residues of micrometric particles of activated charcoal 
coated with water-ice (a type of intimate mixture they refer to as 
"intra-mixture") possessed phase curves that closely resembled the 
phase curve of the Apis/Imhotep border, measured during the low-altitude 
low-phase angle flyby of 67P's nucleus \citep{Feller_2016}.\\
Moreover through the modeling of the scattering properties of the dust 
particles from the nucleus and the inner coma, \citet{Marschall_2020} showed 
that the dust phase function of the inner coma was best-fitted by particles 
ranging in the tens of micrometres, while the phase function of the nucleus 
could be fitted with a model of larger particle aggregates ($>$ 100 $\mu$m).\\ 

As specified in \citet{Lethuillier_2022}, the CoPhyLab dust mixture 
was also designed to replicate the range of particle sizes found for 
67P/C-G. On the one hand, the chosen SiO$_{2}$ material is composed of 
sub-micrometric and micrometric particles, as mentioned above in Sec. 
\ref{sec:material_instruments}. In particular, for this industrially 
manufactured material, \citet{Kothe_2013} reported a mean particle-size 
of 0.63 $\mu$m by count, and of 2.05 $\mu$m by mass. Such sizes are 
consistent with the SEM images of our sample (see Fig. \ref{fig:sem_sio2}). 
Moreover, this material has also a propensity to form aggregates sizing up 
to the millimetre.\\
On the other hand, our juniper charcoal material was grinded then sieved 
with a 50 \mum mesh and \citet{Lethuillier_2022}, resolving optically 
juniper charcoal particles suspended in an ethanol solution, found an average 
particle-size of $\sim$ 24 $\mu$m.\\

As a cross-verification of the characterization of the materials, we 
considered the spectral mixing relation given in \citet{Hapke_1993} 
between physical quantities and single-scattering albedoes for intimate 
mixtures of closely-packed particulates. Assuming that the i$^{th}$ group 
of particles from the mixture contains N$_{i}$ particles, of cross-section 
$\sigma_{i}$ associated with the extinction efficiency Q$_{E, i}$, and 
the single-scattering albedo w$_{i}$, \citet{Hapke_1993} states:
\begin{equation}
    w_{mixture} = \left(\sum_i N_{i}\sigma_{i}\cdot w_{i}Q_{E, i}\right)/
                  \left(\sum_j N_{j}\sigma_{j}\cdot Q_{E, j}\right)
    \label{eq:mix_model_base}
\end{equation}
In the particular case of an ideal binary mixture for which all 
particles of either material have a quasi-spheroidal shape with an 
equivalent diameter D$_{i}$ that is larger than wavelength, the 
Fraunhofer diffraction is negligible (i.e. Q$_{E, i}$ $\sim$ 1) 
according to \citet{Bohren_1983}, and \citet{Hapke_1993} simplifies 
Eq. \ref{eq:mix_model_base} as follows:
\begin{equation}
    w_{mixture} = \frac{x\cdot w_{1}+(1-x)\cdot\zeta\cdot w_{2}}{1+\zeta}
    ~~\text{where} ~ \zeta = \frac{M_{2}}{\rho_{2}D_{2}}\cdot\frac{\rho_{1}D_{1}}{M_{1}}
    \label{eq:mix_model_simpl}
\end{equation}
where M$_{i}$ and $\rho_{i}$ are respectively the bulk and the apparent 
particle densities for the i$^{th}$ material, and x the mass fraction of the 
first material.\\

\begin{figure*}
 \begin{center}
   \includegraphics[width=0.66\linewidth]{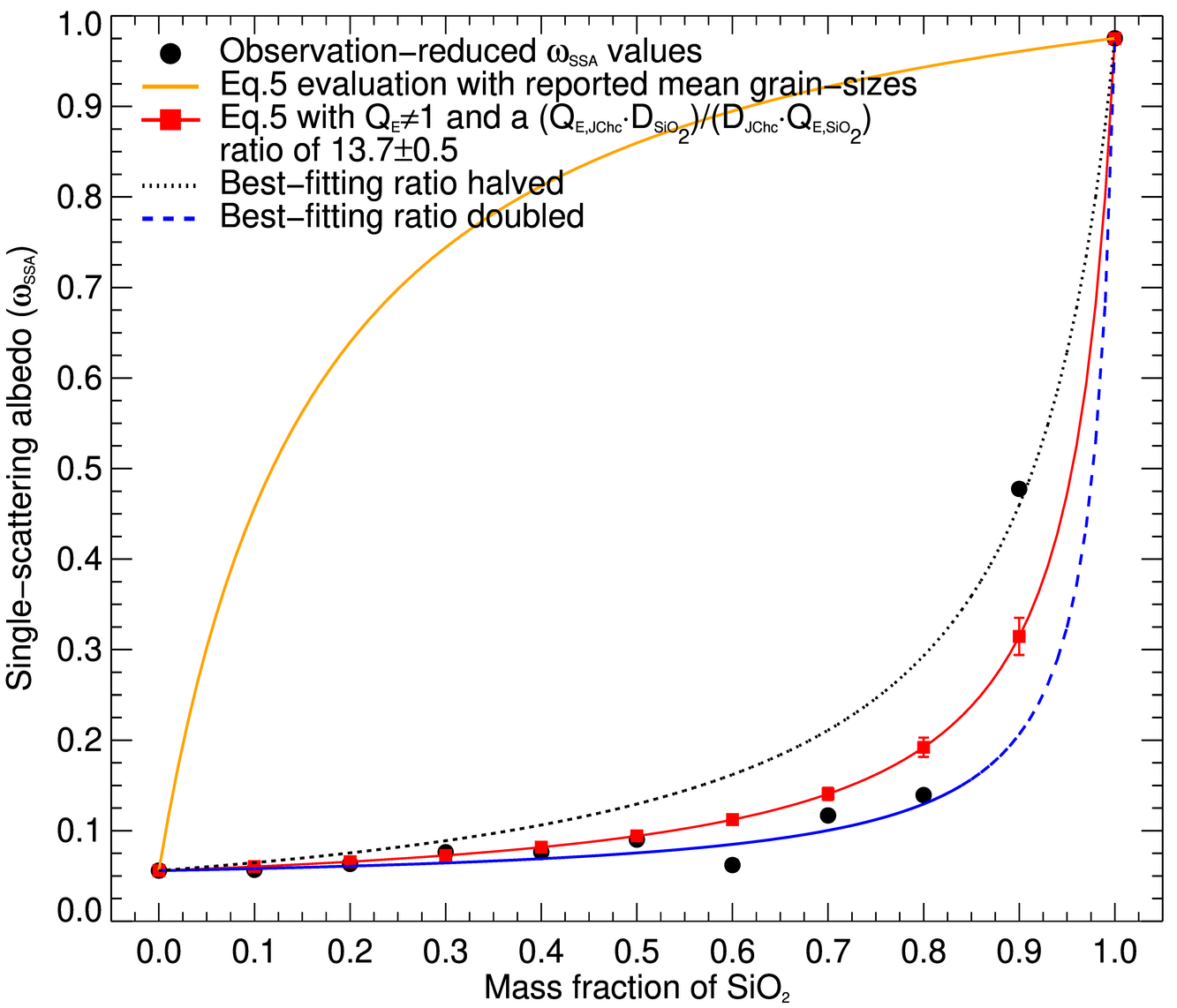}
 \end{center}
  \caption{\label{fig:hpk_intimate} Single-scattering albedo (SSA) values for 
    SiO$_{2}$, juniper charcoal, and the considered intimate mixtures. In this 
    figure, non-visible error-bars are hidden by the plot symbols. This plot 
    highlights the discrepancy between the SSA values derived from the PHIRE2 
    measurements (black circles) and those computed directly with Eq. 
    \ref{eq:mix_model_simpl} using the end-members' reported mean grain-sizes 
    (2.05 \mum for the SiO$_{2}$ and 24 \mum for the JChc) and assuming their 
    particles to be equant (orange line). The divergence of the SSA values for 
    mass fractions 0.6 and 0.9 is further discussed in the main text.}
\end{figure*}

Using that last equation and the values determined previously for either 
material, we plotted the generated curve for all mass fractions against the 
determined single-scattering albedoes for each prepared mixture (respectively 
the orange line and the black points) in Fig. \ref{fig:hpk_intimate}.
We note here that for the benefit of clarity, we only generated the 
said curve using a SiO$_{2}$ mean grain-size of 2.05 \mum, which provided 
the closest result to the SSA values from either reported grain-size.\\
The mismatch between the curve drawn from Eq. \ref{eq:mix_model_simpl} 
and the SSA values is evident. Under the hypotheses upholding this 
equation, the grain-sizes ratio of the two materials would have to 
be inverted for the curve to start matching the SSA values.\\

A close consideration of the end-members' SEM images suggest 
that the discrepancy stems from the hypothesis that the materials' 
particles are equant and each of the same size. While the SiO$_{2}$ 
particles present an ample yet limited particle-size distribution, 
the JChc particles exhibit an wider particle-size distribution 
(from a few hundred of nanometres to hundreds of micrometre) and 
a large diversity of shapes. The presence of long needle-like 
particles (see Fig. \ref{fig_supp:sem_jchc_psd}) is especially of 
note.
Indeed, \citet{Zerull_1976} and \citet{Bohren_1983} demonstrated 
that the scattering properties of non-equant spheroidal particles 
differ significantly from those of spherical particles, and 
the associated simplifications leading to Eq. \ref{eq:mix_model_simpl} 
are therefore unsuited under such premises.\\

However, for the sake of the argument, if one considers 
Eq. \ref{eq:mix_model_simpl} minus the assumption that 
Q$_{E,i}$ = 1 and thus adjusts a
$\frac{Q_{E,JChc}\cdot D_{SiO_{2}}}{D_{JChc}\cdot Q_{E,SiO_{2}}}$ 
ratio with an ordinary least-squares method to fit the SSA values, 
one finds the best-fitting numerical value to be equal to 13.7\mmp 0.5, 
for a quality of fit slope of 0.98±0.02 and a coefficient 
of determination (R$^2$) of 0.99.\\
The associated curve is displayed in Fig. \ref{fig:hpk_intimate} as a 
red line with red squares to highlight the computed values corresponding 
to the mass fractions of the different mixtures.\\
While this curve is well adjusted to mixtures with a lower 
mass-fraction of SiO$_{2}$, it mismatches determined SSA values for 
mixtures with a predominant SiO$_{2}$ fraction, which could have been
interpreted under the premises of Eq. \ref{eq:mix_model_simpl} as 
a difference in terms of grain-sizes in between the mixtures or in 
terms of scatterers.\\

While each investigated sample appears to have a homogeneous 
texture at the millimetre-scale under the naked eye (see Fig. 
\ref{fig_supp:pics}), the SEM images present a different picture.
As mentioned previously, both end-members (Fig. \ref{fig:sem_pure})
present wide grain-size distribution, both materials contain 
sub-micrometric and micrometric particles. In particular, we observe 
individual SiO$_2$ particles that are around one or 
two times the size of the considered wavelength (550 nm) and also find 
individual juniper charcoal particles or their surface elements that 
are $\sim$ 10 times smaller (see Figs. \ref{fig:sem_jchc} (bottom-panel) 
and \ref{fig_supp:sem_jchc_psd}).\\
Thus, both materials contain grains and surface elements, whose 
diameter (or size) is of the order or even smaller than the wavelength 
(550 nm). Such features will likely act as Rayleigh scatterers and 
absorbers, and either contribute to increase or decrease the overall 
reflectance of the material \citep{Hapke_1993}.\\

Moreover as illustrated in Fig. \ref{fig:sem_mixtures}, in 
every mixture investigated with the SEM, we note the presence of large 
and compact SiO$_2$ agglomerates with sizes ranging up to the hundreds 
of micrometers. If any degree of electromagnetic coupling exists between 
particles of such agglomerates, then they would interact with incident 
light similarly to an individual particle of larger size 
\citep{Mustard_1997, Rousseau_2017, Pommerol_2008}.\\

The SEM of the mixtures images present a more complex picture, 
especially those with SiO$_2$ mass fractions higher than 0.5. 
In particular, for the 90\% SiO$_2$/10\% juniper charcoal mixture, 
several larger charcoal particles are visible at the surface, some of 
them appearing as inclusions within large SiO$_2$ agglomerates. For 
mixtures with a slightly higher charcoal content, such textural 
heterogeneities are also noticeable with the continuous presence of 
SiO$_2$ aggregates noted previously, but also with some juniper 
charcoal material appearing as loose elements forming a surface 
surrounding or partially covering the SiO$_2$ aggregates.\\

Thus we interpret the mismatch between the results of the spectral 
mixing model and the SSA values to reflect not only that the 
end-members are composed of particles with a diversity of shapes but 
also as a likely consequence of  the combination of micro-scale 
compositional heterogeneities as well as grain-size effects.\\

We note here that recent researches have investigated the scattering 
properties of irregular particles, for instance through the generation 
of 3D particle shapes and the resolution of the T-matrix for these shapes 
(e.g. \citealt{Petrov_2012}, \citealt{Grynko_2018}). This method has been 
considered to investigate the properties of 67P/C-G's coma and nucleus 
particles.  With equant irregular rough agglomerates of sizes ranging from 
5 $\mu$m to 100 $\mu$m, and composed of an intimate mixture of strongly 
opaque nanometre-sized particles and weakly absorbent particles larger 
in size (up to the micrometre), \citealt{Markkanen_2018} matched phase 
curves of the nucleus and the coma of 67P/Churyumov-Gerasimenko. These 
results underline the usefulness of materials with an extended 
size-distribution. Moreover, and together with the results presented in 
this study, they urge for further investigation of non-homogeneous 
agglomerates composed of particles of varied shapes and sizes.

\newpage

\subsection{Colours and spectral slopes: comparison with small bodies of the 
Solar System}

\subsubsection{The case of 67P/Churyumov-Gerasimenko: spectral slopes and colours}
The nucleus of comet 67P/CG is to date the most extensively characterised 
cometary surface. In particular, although the spectrophotometric properties 
of the nucleus were found to be overall homogenous across both of its lobes, 
its surface displayed a definite diversity of spectral slopes and albedos 
correlated to varied compositional and morphological features, e.g. 
\citet{Barucci_2016}, \citet{Ciarniello_2016}, \citet{Filacchione_2016}, 
\citet{Fornasier_2016}, \citet{Hasselmann_2016}, \citet{Oklay_2016}, 
\citet{Ferrari_2018}, \citet{Feller_2019}, \citet{Fornasier_2019a}, and 
\citet{Hoang_2020}.\\
Among other results, these studies notably established that three 
spectrophotometricaly different types of terrains could be identified across 
the nucleus' surface: those with a shallow spectral slope (S$_{882 nm-535 nm}$ 
$\simeq$ 11--14 \%/100 nm at $\sim$ 50\odeg, using a revised definition of 
\citet{Delsanti_2001}), those with a spectral slope between 14 \%/100 nm and 
18 \%/100 nm, and finally those with a spectral slope above 18 \%/100 nm, at 
$\sim$ 50\odeg (see \citealt{Fornasier_2016} for a discussion on the 67P/CG's 
nucleus phase reddening). While the first group of terrains was correlated 
with surfaces and locations presenting an enhanced water-ice content, the 
third one was associated with water-ice depleted and macroscopically rough 
surfaces (e.g. boulder fields, escarpment ridges). Finally, the second group 
of terrains stood out as the typical dust-covered macroscopically smooth 
expanses present on both lobes of the nucleus.\\

Using the definition of \citet{Fornasier_2015} and the OSIRIS/NAC filter 
profiles\footnote{Available at the Spanish virtual observatory: 
http://svo2.cab.inta-csic.es/index.php and described in \citet{Tubiana_2015}}, 
the corresponding spectral slopes for the CoPhyLab materials and mixtures were 
estimated for a phase angle of $\sim$ 0\odeg and gathered in Table 
\ref{tbl:spcslp}). These values maintain the previously noted trend of a 535 
nm -- 882 nm spectral slope increase with an increasing amount of juniper 
charcoal content, as well as the observation that overall these visible 
spectral slopes are weaker than that of 67P/CG. However, assuming the two 
extreme values found for the phase reddening slope of the 67P/C-G nucleus 
from either observations acquired around August 2014 or around April-August 
2015 (\citealt{Fornasier_2015}; \citealt{Fornasier_2016} supl. mat.), the 
spectral slope offset would be either of $\sim$ 2. \%/100 nm or of $\sim$ 5. 
\%/100 nm.\\
Hence, should any compositional differences be set aside, then the spectral 
slope values of the CoPhyLab mixtures are in appearance consistent with the 
spectral slopes of the water-ice enriched terrains observed on the surface of 
comet 67P/CG. However, if the slope offset associated with the phase reddening 
phenomenon is included, spectral slopes of mixtures with a relative juniper 
charcoal content above 80\% are then comparable with those measured for the 
dust covered areas such as Anuket or Ma'at on 67P/CG's nucleus
\cite{El-Maarry_2015}.\\
\begin{table*}
  \begin{center}
  \caption{\label{tbl:spcslp} Visible spectral slope of the pure materials and 
           their mixtures using the same formula from \citet{Fornasier_2015}, 
           and BVRI colours computed using the canonical Bessell filters. The 
           uncertainties on the charcoal content for the mixtures were 
           evaluated during the sample preparation process. Both analogue or 
           corresponding values available for cometary nuclei flown-by are also 
           listed here for reference.}
  \begin{tabular}{c|c|c|c|c}
    \hline
    Sample             &535--882 nm slope   & B-V            & V-R          & V-I              \\
    (charcoal content) &(\%/100 nm, @0\odeg)& (mag)          & (mag)        & (mag)            \\
    \hline                                 
    0\% (Pure SiO$_2$) & $~$2.8\mmp0.2   &    0.75\mmp0.03   & 0.41\mmp0.01 &    0.77\mmp0.03  \\
    10.0\mmp0.02\%     &   -1.4\mmp0.2   &    0.61\mmp0.03   & 0.34\mmp0.02 &    0.65\mmp0.03  \\
    20.0\mmp0.03\%     & $~$1.0\mmp0.3   & $~$ 0.6\mmp0.1$~$ & 0.35\mmp0.04 & $~$ 0.7\mmp0.1$~$\\
    30.0\mmp0.05\%     & $~$3.8\mmp0.9   &    0.63\mmp0.06   & 0.39\mmp0.01 &    0.78\mmp0.06  \\
    40.0\mmp0.07\%     & $~$5.0\mmp1.$~$ &    0.63\mmp0.1$~$ & 0.40\mmp0.05 & $~$ 0.8\mmp0.1$~$\\
    50.0\mmp0.09\%     & $~$4.0\mmp1.$~$ & $~$ 0.6\mmp0.2$~$ & 0.38\mmp0.08 & $~$ 0.8\mmp0.3$~$\\
    60.0\mmp0.1\%$~$   & $~$5.0\mmp1.$~$ &    0.62\mmp0.04   & 0.39\mmp0.03 &    0.80\mmp0.04  \\
    70.0\mmp0.1\%$~$   & $~$5.0\mmp1.$~$ & $~$ 0.6\mmp0.1$~$ & 0.39\mmp0.02 & $~$ 0.8\mmp0.1$~$\\
    80.0\mmp0.1\%$~$   &   11.0\mmp3.$~$ & $~$ 0.7\mmp0.1$~$ & 0.46\mmp0.06 & $~$ 0.9\mmp0.1$~$\\
    90.0\mmp0.2\%$~$   & $~$7.0\mmp2.$~$ &    0.64\mmp0.07   & 0.41\mmp0.01 &    0.85\mmp0.07  \\
    100\% (Pure JChc)  &   10.0\mmp3.$~$ & $~$ 0.6\mmp0.1$~$ & 0.42\mmp0.07 & $~$ 0.9\mmp0.1$~$\\
    \hline
    \hline
    1P/Halley$^{1}$      & [17.4\mmp0.3] &    0.72\mmp0.04   & 0.41\mmp0.03 &    0.80\mmp0.07  \\
    9P/Tempel 1$^{2}$    & [12\mmp 1]    &    0.84\mmp0.01   & 0.50\mmp0.01 &    0.99\mmp0.02  \\
    67P/CG$^{3}$         & [11\mmp 2]    &    0.83\mmp0.08   & 0.54\mmp0.05 &    1.00\mmp0.07  \\
    67P/CG$^{4}$         & [   20   ]    &    0.73\mmp0.07   & 0.57\mmp0.03 &    1.06\mmp0.05  \\
    103P/Hartley 2$^{5}$ & [ 7\mmp 3]    &    0.75\mmp0.05   & 0.43\mmp0.04 &    0.84\mmp0.05  \\
    \hline
  \end{tabular}
  \end{center}
  \begin{flushleft}
    Notes: 1: \citet{Thomas_1989,Lamy_2004}, this spectral slope is 
    estimated there between 440 nm and 813 nm at small phase angles. -- 2: 
    \citet{Li_2007}, this spectral slope was estimated there between 310 nm 
    and 950 nm at 63\odeg of phase angle. -- 3: \cite{Tubiana_2008, 
    Tubiana_2011}, this spectral slope was estimated between 436 nm and 797 
    nm at phase angles lower than 10\odeg. This particular set of values was 
    derived from ground-based observations at heliocentric distances larger 
    than 4.5 au, while no coma features were detected. -- 4: 
    \cite{Ciarniello_2015}, this spectral slope was estimated between 550 nm 
    and 800 nm. -- 5: \citet{Li_2013}, this spectral slope was computed 
    between 400 nm and 850 nm, at 85\odeg of phase angle.
  \end{flushleft}
\end{table*}
Although this phase reddening phenomenon has been observed for the Moon' 
surfaces \citep{Gehrels_1964}, and those of asteroids and meteorites 
(e.g. \citealt{Sanchez_2012, Fornasier_2020} and references therein), the 
mechanisms behind phase reddening are still the subject of studies. This 
linear increase of spectral slopes with the phase angle has been associated 
with physical properties such as particle single-scattering, and sub-micron 
roughness of the regolith \cite{Grynko_2008, Schroeder_2014, Ciarniello_2020}.
In the case of comet 67P/C-G, \citet{Fornasier_2016} further associated 
variations of the phase reddening with compositional changes of the surface 
as the comet passed through perihelion.

\subsubsection{The case of 67P/Churyumov-Gerasimenko: a look at the spectrum}
\begin{figure*}
  \begin{minipage}[h]{0.47\textwidth}
  \begin{center}
  \includegraphics[width=\linewidth]{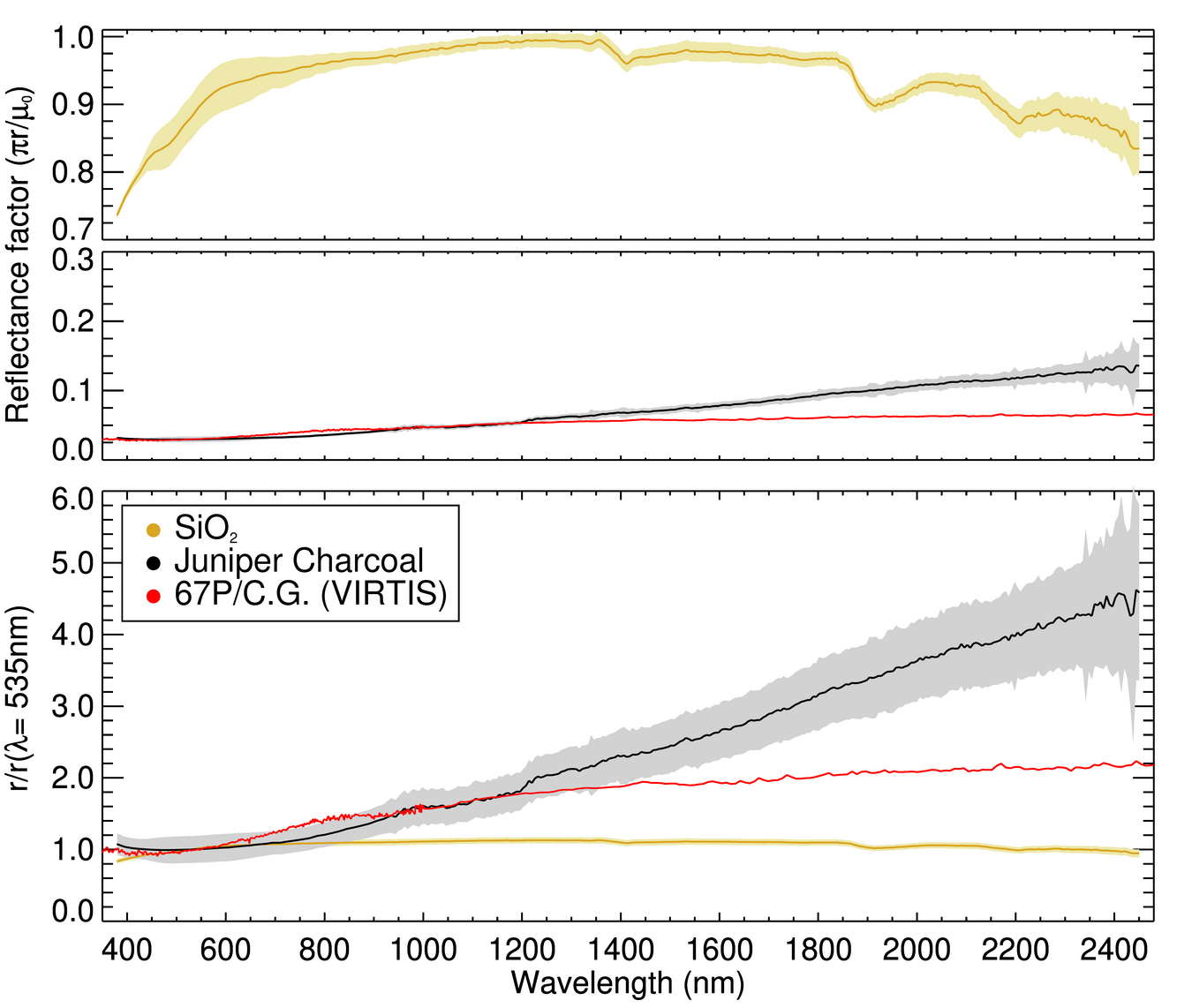}
  \end{center}
  \end{minipage}
  \hfill
  \begin{minipage}[h]{0.47\textwidth}
  \begin{center}
  \includegraphics[width=\linewidth]{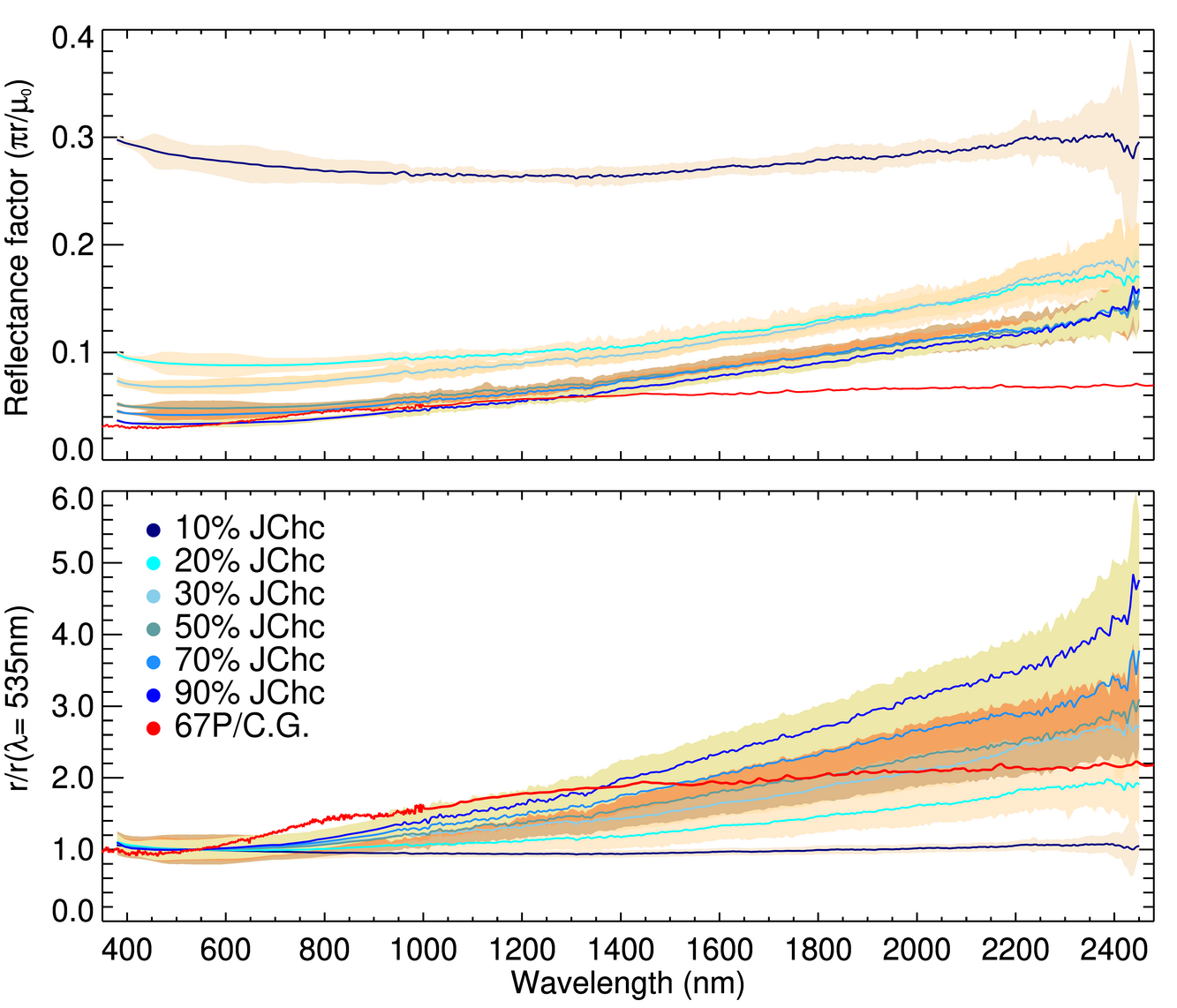}
  \end{center}
  \end{minipage}
  \caption{\label{fig:spectra_comparison} Spectra of the investigated 
           end-members and associated mixtures plotted alongside the spectrum 
           from the Aswan terrace on 67P/C-G (adapted from \citealt{Capaccioni_2015}, 
           and depicted by the solid red curves in this figure). As in Fig. 
           \ref{fig:spectra_mix}, the color filled envelope around the solid 
           curves correspond to the associated dispersion, and only the spectra 
           for the 10\%, 20\%, 30\%, 50\%, 70\% and 90\% juniper charcoal 
           mixtures are displayed for clarity. -- 
           Top plots: REFF spectra for the end-members (left) and for the 
           associated intimate mixtures (right). -- 
           Bottom plots: Associated reflectance spectra normalised at 535 nm 
           the end-members (left) and for the associated intimate mixtures 
           (right).
          }      
\end{figure*}
Beyond the comparison of colours and spectral slopes in the visible 
domain, we sought to compare our VIS-NIR spectra with one from a typical 
surface of 67P/C-G's nucleus. We plot in Fig. \ref{fig:spectra_comparison} 
the spectrum from part of the dust-covered terrace overhanging the Aswan 
cliff near the neck of the comet \citep{Capaccioni_2015, Pajola_2016}.\\
For the purpose of this study, we convert this spectrum given in radiance factor (RADF) in \citet{Capaccioni_2015} to reflectance factor (REFF). Using the reconstructed trajectory of Rosetta at the moment of the observation \citep{Acton_2018} and a shape 
model of the comet \citep{Preusker_2017}, we determined that the incidence, 
emergence and phase angles were respectively of 57\mmp4\odeg, 86\mmp4\odeg, 
and 28.95\mmp 0.03\odeg for this part of the terrace.\\

As discussed in \citet{Capaccioni_2015}, this spectrum does not 
exhibit specific spectral features across the 380 -- 2500 nm range and 
presents a steady low reflectance (from $\sim$ 0.03 at 380 nm and 535 nm to 
$\sim$ 0.07 at 2500 nm). Compared to the SiO$_2$ spectrum, both the juniper 
charcoal and this 67P/C-G spectrum are strikingly darker. However, although 
these two spectra exhibit comparable low REFF values across the 380 -- 1200 
nm interval, the juniper charcoal' spectrum keeps on increasing toward 
higher wavelengths, reaching up to almost twice the REFF values of 67P/C-G's 
beyond 2400 nm (see Fig. \ref{fig:spectra_comparison}, top-left). This 
difference of behaviour in the near-infrared domain is particularly evident 
in the normalised reflectance plot (Fig. \ref{fig:spectra_comparison} 
bottom-left), across which 67P/C-G' normalised reflectance values are 
plateauing toward 2.1. On the other hand, the JChc normalised spectrum 
increases steadily from $\sim$ 1.8 at 1200 nm to $\sim$ 4.6 at 2400 mm.\\
This plot also casts light on the differences between 67P/C-G's and JChc'
spectra in the visible domain, and in particular across the 535 nm to 800 nm 
interval over which the spectral slope for 67P/C-G is of 12 \%/ 100 nm, 
whereas the corresponding slope for the juniper charcoal is of $\sim$ 6\mmp 5 
\%/ 100 nm. This difference in behaviour extends beyond the visible domain
with the noted steady increase of the JChc' relative reflectance, while the 
67P/C-G' relative reflectance spectrum inflects past $\sim$ 800 nm, and 
then slowly increases from $\sim$ 1.5 to $\sim$ 2.1 (reached at 2450 nm).\\
We note here that as the 67P/C-G' spectrum inflects towards this 
slow-raising segment, together with the juniper spectrum, both spectra 
present comparable reflectance factor values across the 1000 -- 1200 nm 
interval. Hence, although the 67P/C-G and juniper charcoal spectra exhibit 
overall different behaviours, across the 380 -- 535 nm and 1000 -- 1200 nm 
intervals, the juniper charcoal REFF spectrum presents comparable values to 
those of the 67P/C-G' REFF spectrum of a dust-covered area of the nucleus' 
surface.\\

While transitioning from a SiO$_2$-fraction dominated mixture to a 
juniper charcoal fraction dominant one, the spectra of these intimate 
mixtures show shapes that differ significantly from that of 67P/C-G's 
nucleus. Both the reflectance factor and the relative reflectance plots of 
Fig. \ref{fig:spectra_comparison} (right column) illustrate this mismatch 
across the 400 nm -- 2450 nm range. The REFF plot (Fig. 
\ref{fig:spectra_comparison}, upper-right) highlights that only spectra of 
mixtures with more than 50\% of juniper charcoal partially match or have 
reflectance factor values close to those of this 67P/C-G' spectrum, over 
the 400 nm -- 1300 nm interval. However, similarly to the pure juniper 
charcoal spectrum, these same spectra exhibit a continuous increase of 
their reflectance above 1400 mm, while this 67P/C-G spectrum does not reach 
REFF values above 0.075.\\
Additionally, although the relative reflectance plot (Fig. 
\ref{fig:spectra_comparison}, lower-right) highlights that all spectra of 
mixtures with more than 20\% of juniper charcoal by mass each share one 
common ratio with the normalised spectrum of 67P/C-G in the near-infrared 
range, none of the investigated mixtures reproduce a spectrum with the 
right curvatures and a neutral-to-moderate slope across the near-infrared 
domain.
\subsubsection{Comparison with other small bodies}
\begin{figure*}
 \begin{center}
 \includegraphics[width=0.50\linewidth]{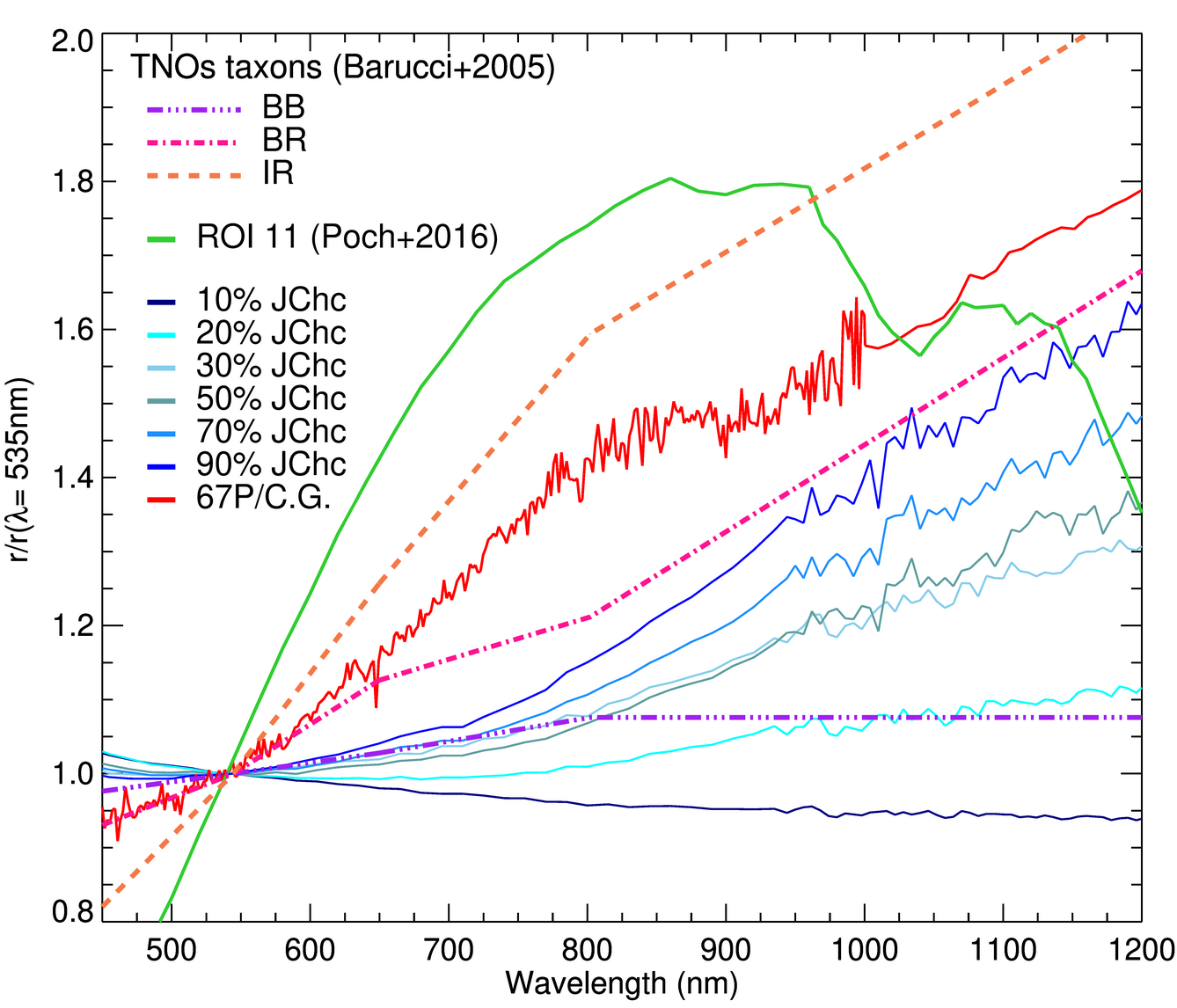}
 \end{center}
  \caption{\label{fig:comparison_tnos} Comparison of the CoPhyLab mixtures' 
    spectra with that of 67P/CG and the normalised reflectances associated 
    with the colors of the TNOs taxons discussed in \citet{Barucci_2005}.
    Although all spectra were photometricaly corrected, it should be noted 
    that the CoPhyLab spectra were acquired at $\sim$ 0\odeg of incidence 
    and phase angles, while the area, from which the spectrum from 67P/CG's 
    “body” was measured, was observed under $\sim$57\mmp 3\odeg of incidence, 
    $\sim$86\mmp 2\odeg of emergence and $\sim$ 28.94\mmp 0.05\odeg of phase 
    angle. The 1-$\sigma$ error-bars are omitted for clarity.}
\end{figure*}

Trans-Neptunian Objects (TNOs) are generaly assumed to form one of the 
reservoirs of Jupiter family comets (JFCs), such as comet 67P/C-G\footnote{We 
note here however the dynamical history of 67P/C-G can not be asserted before 
1923 \citep{Maquet_2015}.} \citep{Gladman_2008}. Thus, Fig. 
\ref{fig:comparison_tnos} illustrates the comparison of the CoPhyLab mixtures 
to a subset of this larger group of bodies, considered in \citet{Barucci_2005}, 
who applied a clustering algorithm to identify classes of TNOs based on their 
spectral properties across the VIS-NIR domain.\\

The TNOs taxons, labeled in Fig. \ref{fig:comparison_tnos} as BB, BR, and 
IR, respectively refer to the TNOs with small or no colour changes (i.e. 
displaying a “blue” spectral behaviour) across the considered wavelength 
domain, those with a slight-to-moderate colour change (i.e. displaying a 
“blue-red” spectral behaviour), and those displaying moderate-to-strong 
colour changes (i.e. presenting an moderately “red“ spectral behaviour).\\
As discussed in \citet{Fornasier_2015} and illustrated here in Fig. 
\ref{fig:comparison_tnos} with the 67P/C-G spectrum from 
\citet{Filacchione_2016}, the normalised reflectance of 67P/CG's typical 
terrains remains in between the “blue-red” and “red” profiles, whereas the 
illustrated CoPhyLab mixtures exhibit a diversity of spectral behaviours 
ranging from “bluer” than the TNOs' BB group, to a behaviour intermediate 
between the BB and BR groups. This relative diversity of spectral profiles 
underlines the difference previously noted for the 535-882 nm spectral 
slopes (see Table \ref{tbl:spcslp}). Through their monotonically 
increasing and incurved profile, the spectra of the CoPhyLab mixtures are 
also distinctly different from spectra of the water-ice/red 
tholins/activated charcoal intimate mixtures considered in 
\citet{Poch_2016a}, which exhibited conspicuously higher spectral slopes 
across the visible range (see Fig. \ref{fig:comparison_tnos} and 
\citealt{Feller_2016}), or from the PSOC-1532/pyrrhotite/dunite intimate 
mixture of sub-micron grains investigated by \citet{Rousseau_2017} (see Fig. 
10 of that paper), which also displayed visible spectral slopes larger than 
that of 67P/CG yet combined with an almost flat spectrum across the NIR 
domain.\\

\begin{figure*}
 \begin{center}
   \includegraphics[width=0.83\linewidth]{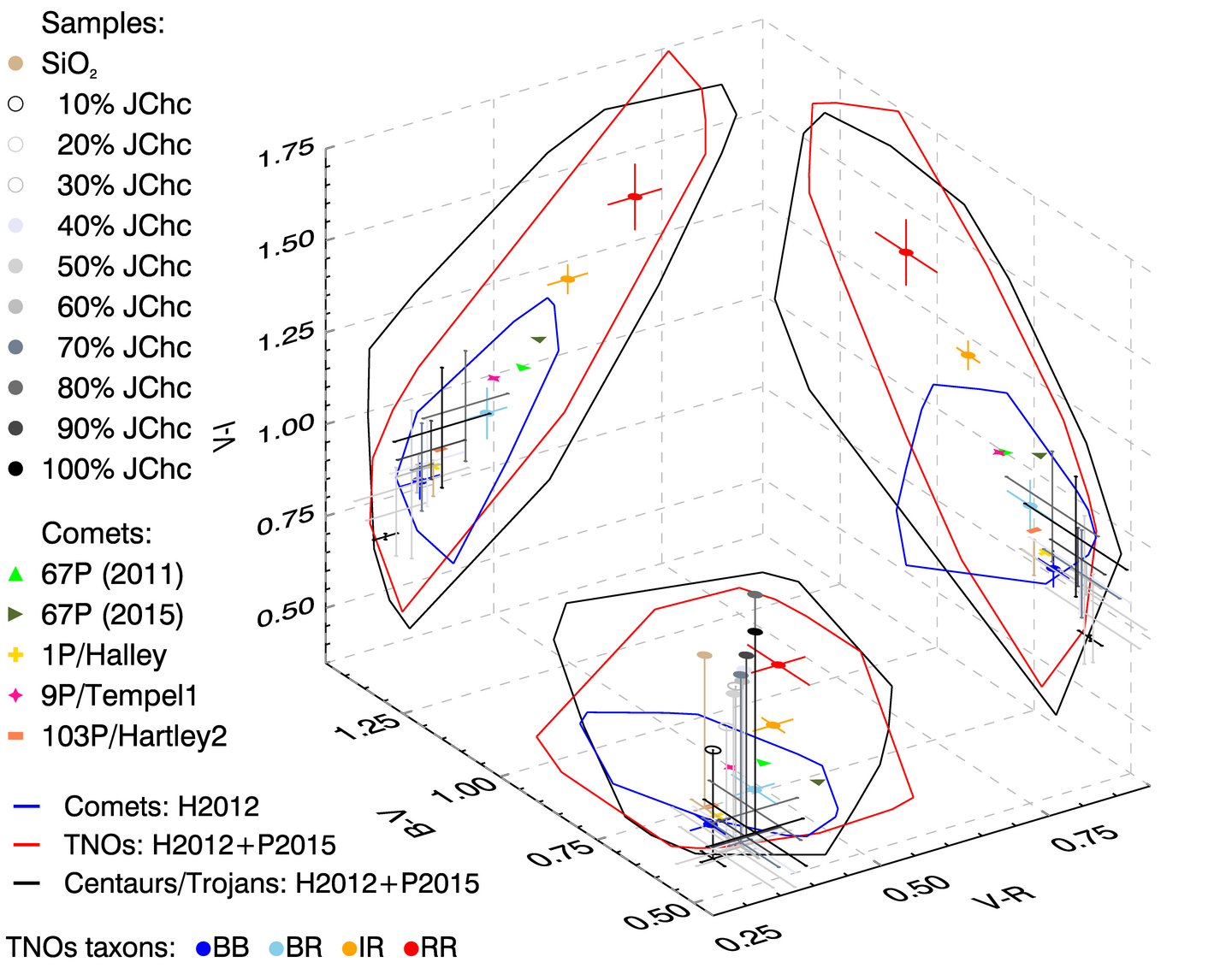}
 \end{center}
 \caption{\label{fig:phot_bessell} Diagram of the standard BVRI color differences 
    for the CoPhyLab samples compared to those of comets visited by spacecrafts 
    (see Table \ref{tbl:spcslp}), and to other small bodies of the Outer 
    Solar System. The area defined by the groups Comets, TNOs and Centaurs/Trojans 
    were derived from the catalogues of \citet{Hainaut_2012} (H2012) and 
    \citet{Peixinho_2015} (P2015), while the TNOs taxons correspond to those 
    defined by \citet{Barucci_2005}.}
\end{figure*}

Furthering the comparison to other small bodies beyond the main-belt, the 
color indices of the CoPhyLab mixtures were computed using standard 
BVRI filter profiles\footnote{Also available at the aforementioned VO 
\href{http://svo2.cab.inta-csic.es/svo/theory/fps3/index.php?mode=browse&gname=Generic&asttype=}{(SVO Bessell filters)}}
\citep{Bessell_1990} and then plotted against the colors indices of comets, 
Centaurs and Trojans, as well as TNOs collected and presented in 
\citet{Hainaut_2012} and \citet{Peixinho_2015}. The authors of these studies 
gathered the (B-V), (V-I) and (V-R) indices for each of 24 comets, 144 Centaurs 
and Trojans, as well as 195 TNOs. The convex hulls for each group of objects 
are plotted in Fig. \ref{fig:phot_bessell}, alongside the colours indices of 
each TNO taxon defined in \citet{Barucci_2005}, and plotted here for reference, 
and those available for comets visited by spacecrafts (see Table 
\ref{tbl:spcslp}).\\

As illustrated in Fig. \ref{fig:phot_bessell}, although the B-V and V-I colour 
indices of the CoPhyLab mixtures place inside, or close to, the V-R/V-I and 
B-V/V-I projections of the comets group's hull (with the exception of the 10\% 
juniper Charcoal mixture), a closer look at the V-R/B-V projections reveals 
their actual positions with respect to the different groupings. 
Almost all CoPhyLab materials exhibit overall colour indices consistent with 
the Centaurs and Trojans grouping, with the exception of the 10\% and 20\% 
JChc mixtures, whose B-V and V-R indices place them just on the outside of 
the hull. Similarly, these two indices also set all the CoPhyLab mixtures 
containing between 30\% and 70\% of juniper charcoal by mass on the lower edge 
of the TNOs group's hull. On the other hand, only the pure juniper Charcoal 
sample and the mixture containing 90\% juniper Charcoal are firmly set within 
the TNOs' hull, with their respective propagated errors encroaching the volume 
of colour indices of comets.\\
Lastly, only the pure sample of SiO$_{2}$ and the measured spectrum of the 
mixture with 80 wt. \% juniper charcoal present colour indices consistent 
with the diversity of the colours indices specific to the comets' group. In 
particular, the SiO$_{2}$'s B-V and V-R indices closely match those of 
1P/Halley and 103P/Hartley 2. While the 80\% charcoal mixture differs 
from the other samples by its colors indices, these values remain nevertheless 
within the volume defined by the colours indices and their error-bars of the 
other mixtures.
\FloatBarrier
\section{Conclusions}
In this paper, we have reported the first spectroscopic and photometric 
measurements for juniper charcoal and for intimate mixtures of juniper charcoal 
and silicon dioxide particles. Both series of measurements show that juniper 
charcoal, the dark fraction of the mixture, governs the overall 
spectro-photometric properties of the mixture. \\
While we find that the spectral behaviour of either end-members or mixtures do 
not provide a satisfactory match to an average spectrum of 67P/C-G's dusty 
surfaces, the consideration of the spectral offset associated with the phase 
reddening phenomenon makes the visible spectral slopes of mixtures 
with more than 50 wt. \% of juniper charcoal comparable to those of the 
average dusty surfaces on 67P/C-G's nucleus. Comparison of the color-indices 
associated with the investigated material with those of small bodies of the 
Solar System shows that while some mixtures present color-indices consistent 
with values observed for comets, all the investigated materials fall more 
closely within the range of values expected for the bluest members of the TNOs 
as well as Centaurs and Trojans objects.\\

We also have shown that the photometric measurements of both juniper charcoal 
and silicon dioxide are best modeled as superficially porous and backscattering 
surfaces, presenting a more moderate opposition effect surge than the surfaces 
of 67P/C-G's nucleus. These two materials most strongly differ through their 
single-scattering albedos. This albedo difference further transpires, for 
instance, when computing their respective geometrical albedos at 550 nm, with 
silicon dioxide having a geometric albedo about as bright as a 99 \% 
reflectance calibration target, while that of the juniper charcoal would be, by 
that standard, about 25\% lower than the geometrical albedo of the Imhotep/Ash 
border of the nucleus.\\
Furthermore, we have shown that the progression of the albedos of the intimate 
mixtures can be best fitted by a decreasing exponential law with respect 
to an increase in the juniper charcoal mass fraction, thus highlighting the 
strong influence of the more absorbent fraction over the overall reflectance 
properties of these intimate. The modeling of the variation of the mixtures' 
single-scattering albedos at 550 nm, with respect to the juniper charcoal
content, using a scale relation between the cross-sectional extinction 
efficiencies per unit length of the materials also supports this result. 
Moreover, collected SEM images show that the large distributions in 
scatterer-sizes for either material and the presence of large scale silicon 
dioxide aggregates are consistent with a Q$^{*}$ ratio as large as 13.7\mmp 
0.5.\\

We have thus found that intimate mixtures of juniper charcoal and silicon 
dioxide present some spectroscopic and photometric properties that are 
consistent with those of certain small bodies of the Solar System, such as 
TNOs and Centaurs, and some of these properties are close to, or of the same 
order as those found for the surface of 67P/Churyumov-Gerasimenko's nucleus. 
While certain results presented in this study could warrant further 
investigations in the frame of the future experiments, either to further the 
physical likeness of the mixture to some of the observed dust- or pebble-covered 
surfaces of 67P/Churyumov-Gerasimenko or to investigate its physical properties 
at the micrometre and sub-micron scale, these mixtures will be considered in the 
upcoming sublimation experiments performed with the large CoPhyLab simulation 
chamber.

\section*{Acknowledgements}
This work was carried out in the framework of the CoPhyLab project funded by 
the D-A-CH programme (1620/3-1 and BL 298/26-1 / SNF 200021E 177964 / 
FWF I 3730-N36).\\
Some of this work made use of the NAIF/SPICE library \citep{Acton_2018}, of 
the ESA/ROSETTA reconstructed trajectory kernels provided by ESA.\\
The samples of juniper charcoal used in this study were grounded with the 
equipment of Dr. Eggenberger from the department of geological sciences of 
the University of Bern. 
Electron microscopy sample preparation and imaging were performed with 
devices supported by the Microscopy Imaging Centre (MIC) of the University 
of Bern. 
The CHN elemental analyses mentioned in this work were performed by the team 
of Pr. Dr. Schürch from the department for chemistry, biochemistry and 
pharmacy of the University of Bern 
(\href{https://www.dcbp.unibe.ch/dienstleistungen/analytik_ars/index_ger.html}{www.dcbp.unibe.ch/}).\\
C.F. expresses its most sincere thanks to N. Ligterink, M. Lopez-Antu\~na, L. Patty, and A. Springmann for their continuous 
support and fruitful discussions.
The manuscript benefited from many helpful suggestions and recommendations 
provided by Dr. Stefanus Schroeder in his review of this manuscript.

\section*{Data Availability}
All spectroscopic and photometric data acquired for this study are available 
on the SSHADE database (\hyperlink{https://doi.org/10.26302/SSHADE/EXPERIMENT_CF_20200723_000}{10.26302/SSHADE/CF-1} and \hyperlink{https://doi.org/10.26302/SSHADE/EXPERIMENT_CF_20200813_000}{10.26302/SSHADE/CF-2}). In addition, the entire set of data and characterizations 
including the SEM images of the investigated samples will be made available on the Zenodo plateform (\hyperlink{https://doi.org/10.5281/zenodo.10280355}{10.5281/zenodo.10280355}) upon publication.


\bibliographystyle{mnras}
\bibliography{./BiblioJabRef} 




\appendix
\section{Supplementary material}
\subsection{juniper charcoal particle-size range}
\begin{figure}
 \begin{minipage}[c]{\linewidth}
 \begin{center}
  \includegraphics[width=\linewidth]{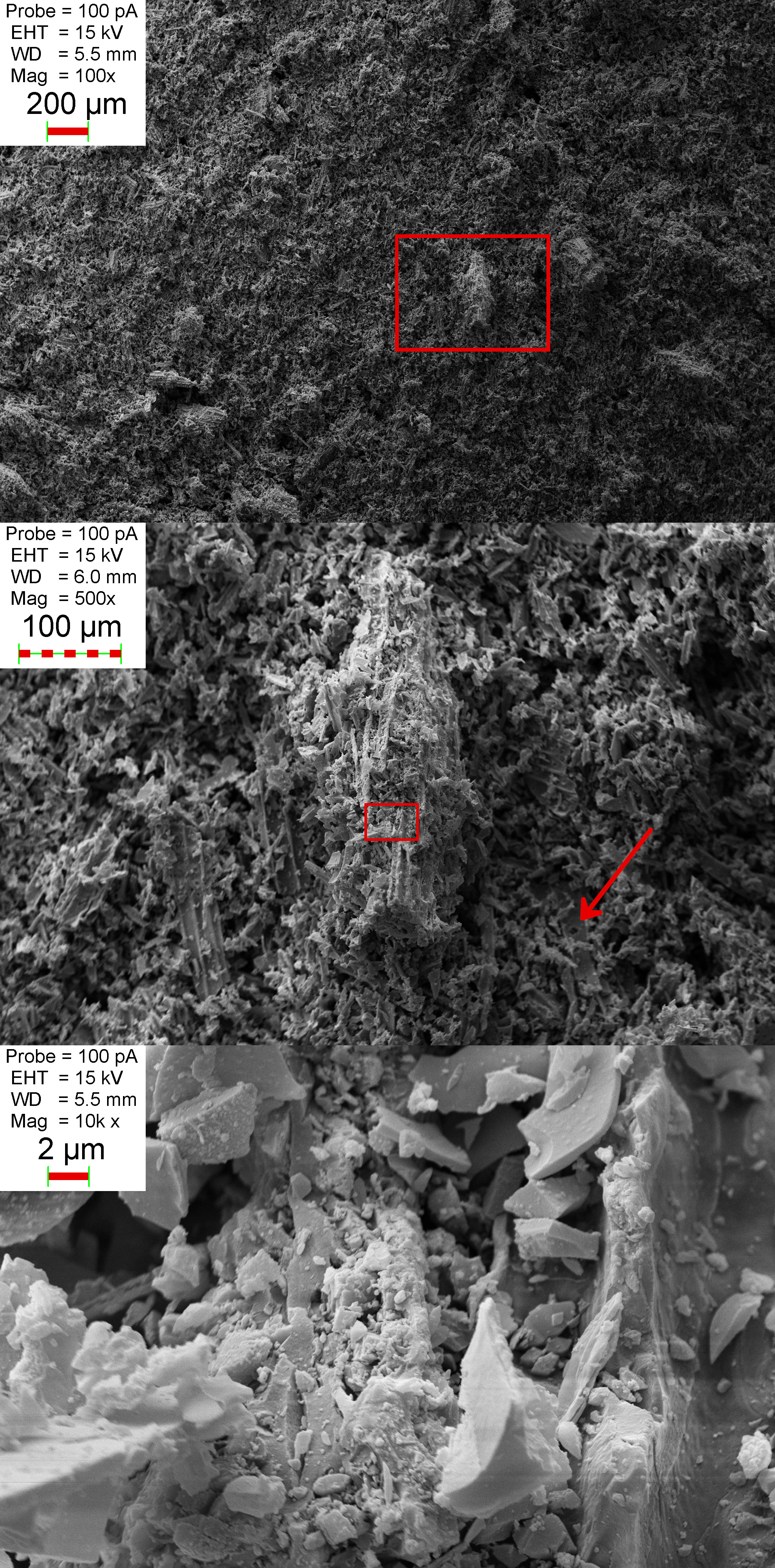}
  \caption{\label{fig_supp:sem_jchc_psd} Alternate zoom on a remnant of the 
      juniper wood structure illustrating the wide range of particle-sizes 
      present in the juniper charcoal sample, which spawns from the tens of 
      nanometers to the hundreds of micrometers. In the middle picture, the 
      red arrow points to a pore in another juniper wood remnant of a vessel 
      wall.}
  \end{center}
 \end{minipage}
\end{figure}
\FloatBarrier

\subsection{Remnant of organic structures within the juniper charcoal}
\begin{figure}
 \begin{center}
 \begin{minipage}[c]{\linewidth}
   \includegraphics[width=\linewidth]{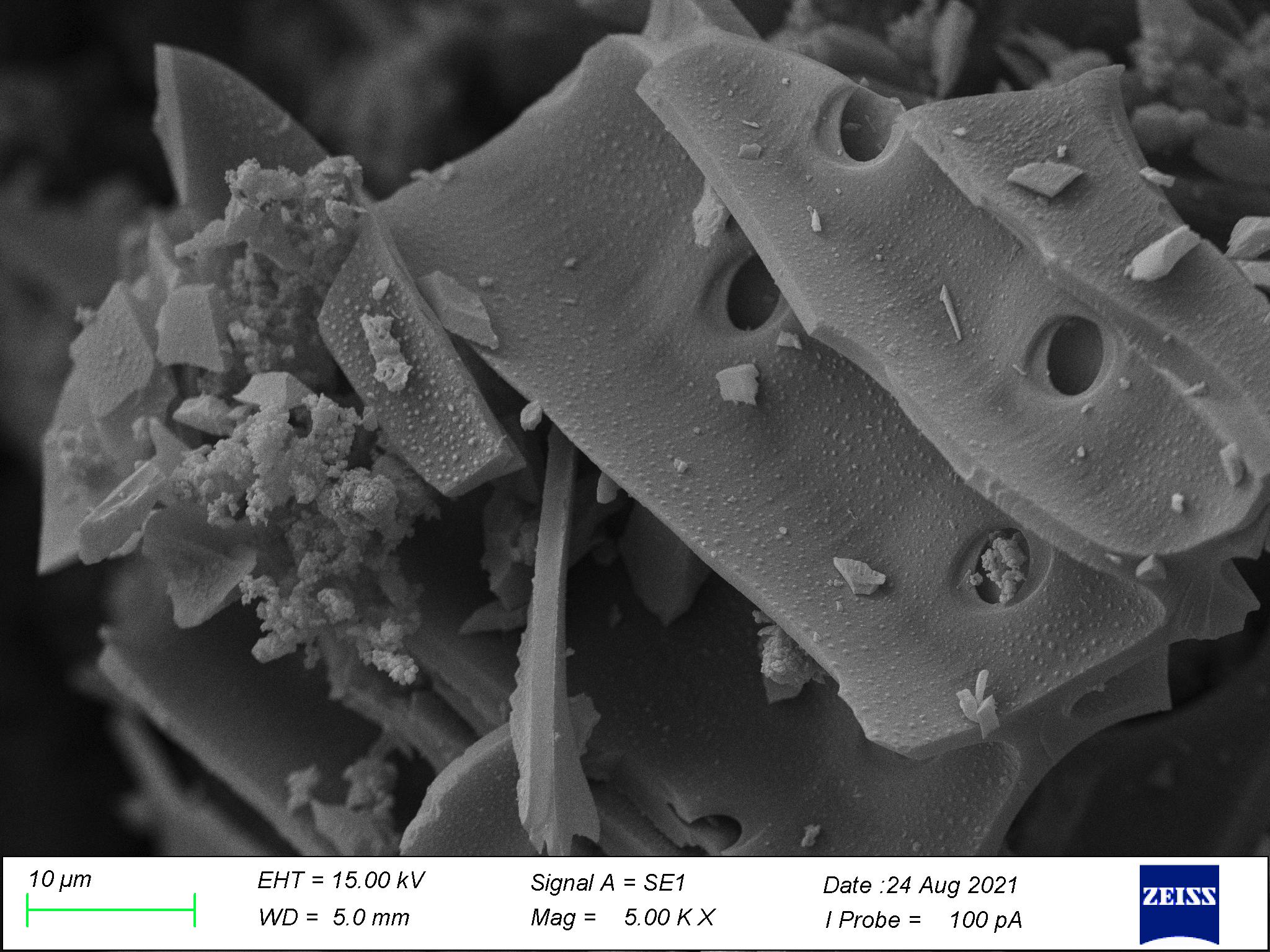}
   \caption{\label{fig_supp:sem_jchc_pores} Zoom-in on the Cophylab mixture 
       with 20\% juniper charcoal, exhibiting small aggregates of SiO$_{2}$ 
       particles sticking to the remnant of a juniper wood vessel structure.
       Micrometre-sized vessel wall pores and lignin protuberances hundreds 
       of nanometres in sized are also clearly distinguishable in this image.}
 \end{minipage}\hfill
 \begin{minipage}[c]{\linewidth}
   \includegraphics[width=\linewidth]{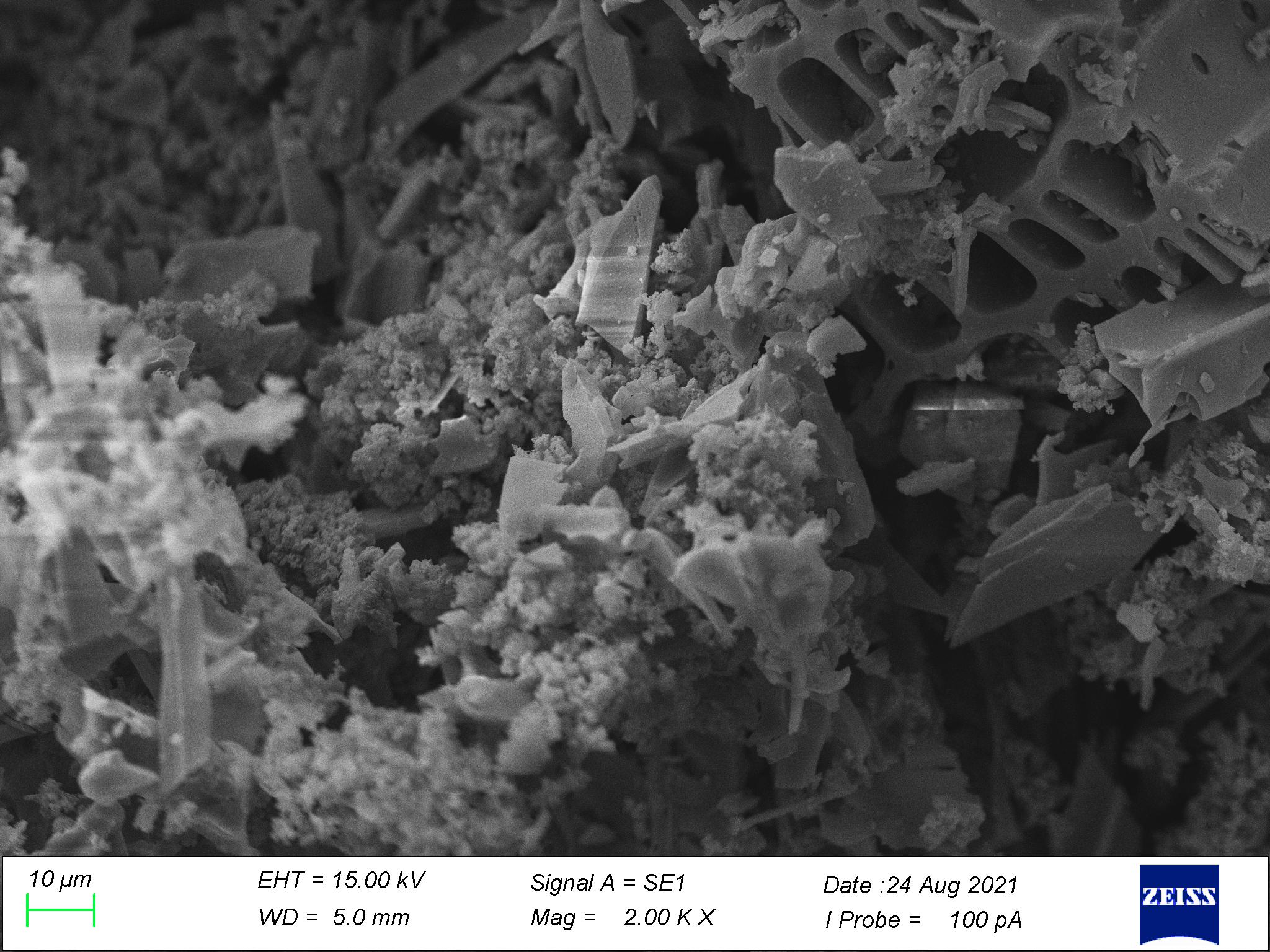}
   \caption{\label{fig_supp:sem_jchc_tubes} Zoom-in on the Cophylab mixture 
       with 20\% juniper charcoal, illustrating the tangle of SiO$_{2}$ grains 
       and juniper charcoal fragments. In the background, a large remnant of 
       juniper wood structure is also visible, which exhibits contiguous 
       micrometre-sized cavities, which are consistent with fibre lumens.}
 \end{minipage}
 \end{center}
\end{figure}
%
\begin{figure*}
 \begin{center}
 \begin{minipage}[t]{0.80\linewidth}
  \includegraphics[width=\linewidth]{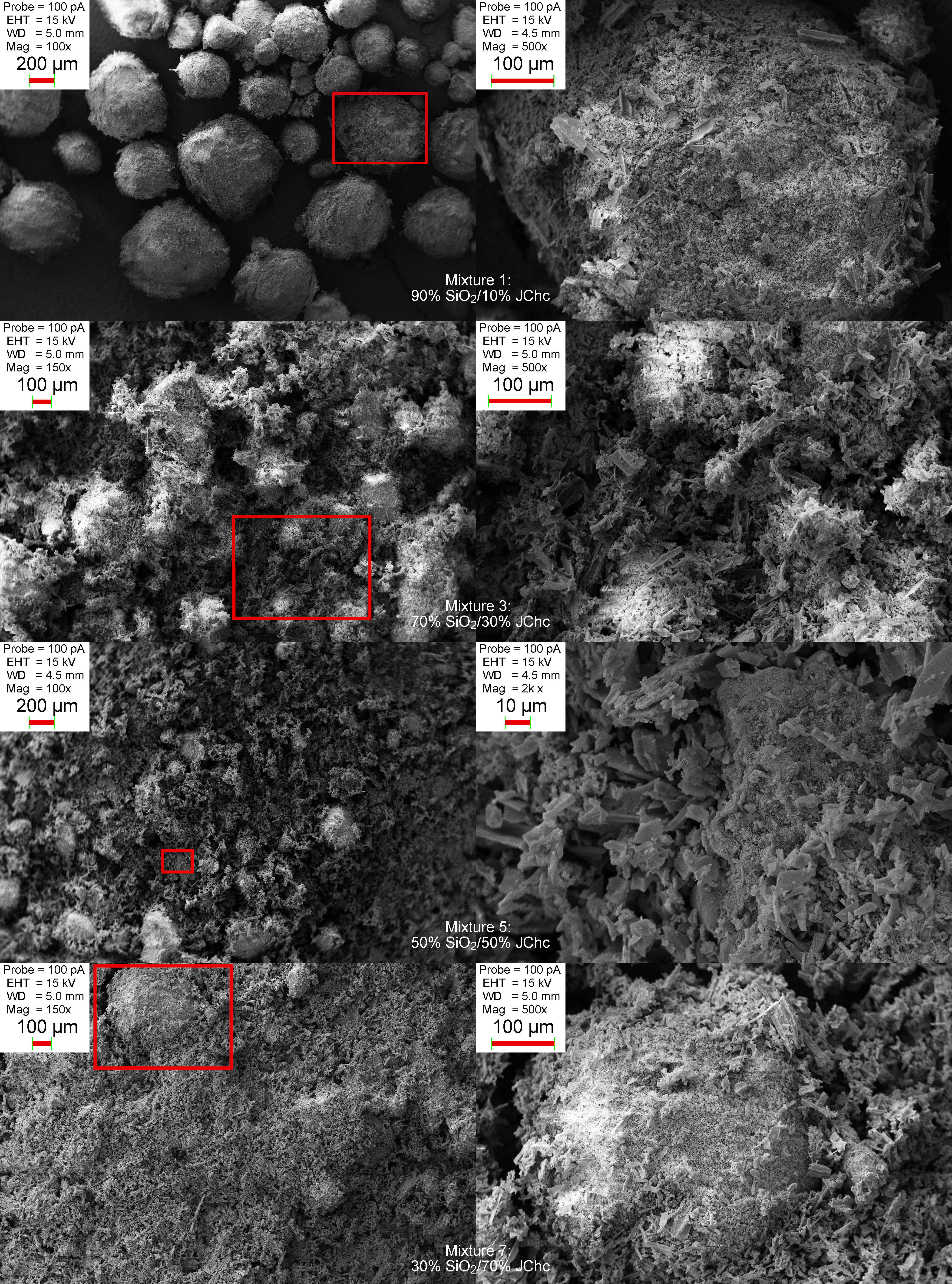}
 \end{minipage}\vfill
 \begin{minipage}[b]{0.80\linewidth}
  \begin{minipage}[c]{0.50\linewidth}
   \includegraphics[width=\linewidth]{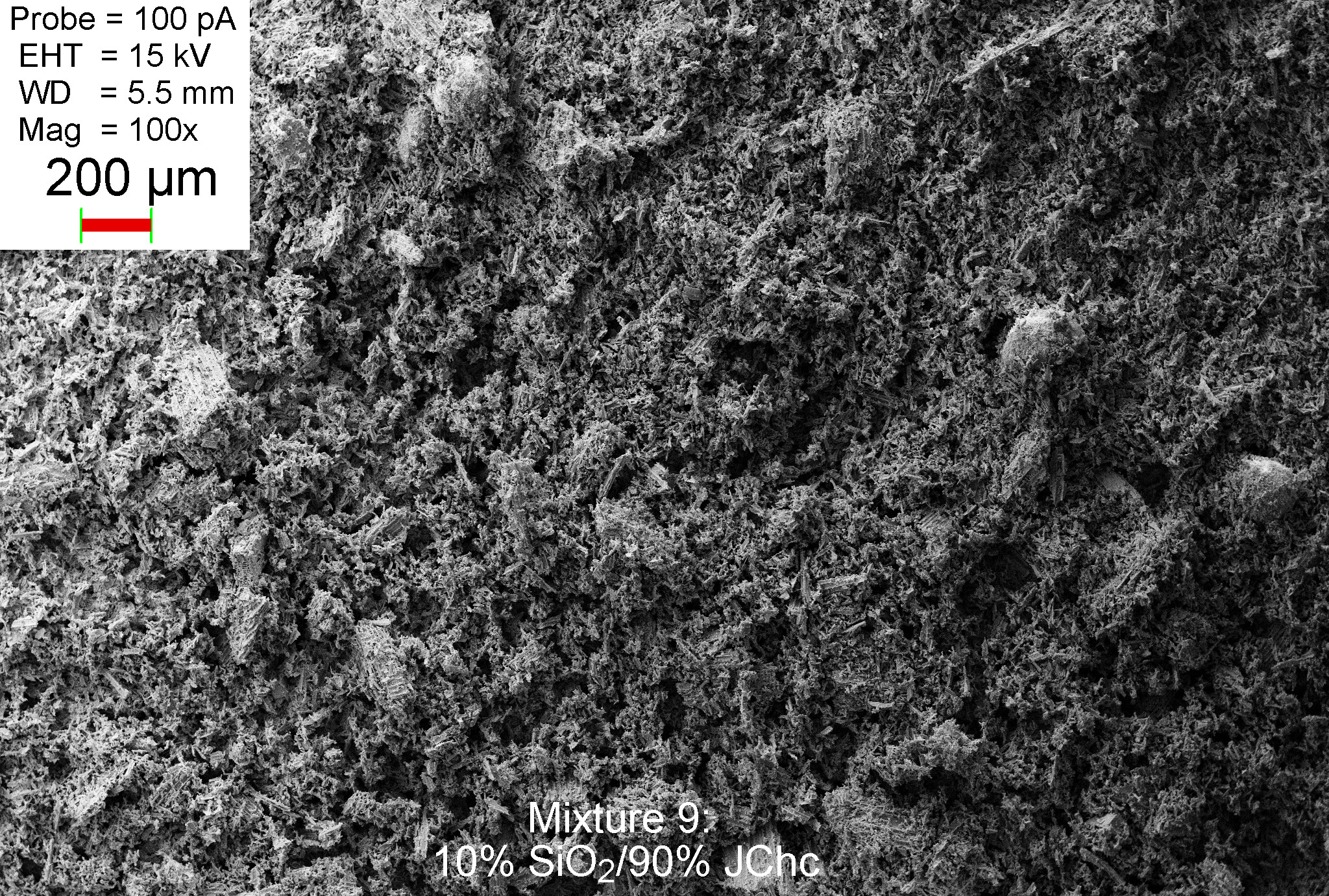}
  \end{minipage}\hfill
  \begin{minipage}[c]{0.49\linewidth}
    \caption{\label{fig:sem_mixtures} Panel of SEM images for the CoPhylab 
      mixtures with 10\%, 30\%, 50\%, 70\%, and 90\% of juniper charcoal by 
      mass. Zoom-in of the areas under the red squares are displayed in the
      SEM images of the right column. \\
      These SEM images notably highlight micro-scale compositional 
      heterogeneities, and reveal the presence of micrometre-sized SiO$_2$ 
      agglomerates for any amount of juniper charcoal, as discussed in the 
      main text.}
  \end{minipage}
 \end{minipage}
 \end{center}
\end{figure*}
\FloatBarrier

\subsection{Pictures of the samples}
 \begin{figure*}
 \begin{center}
  \begin{minipage}[t]{0.90\linewidth}
   \begin{minipage}[l]{0.49\linewidth}
   \includegraphics[width=\linewidth]{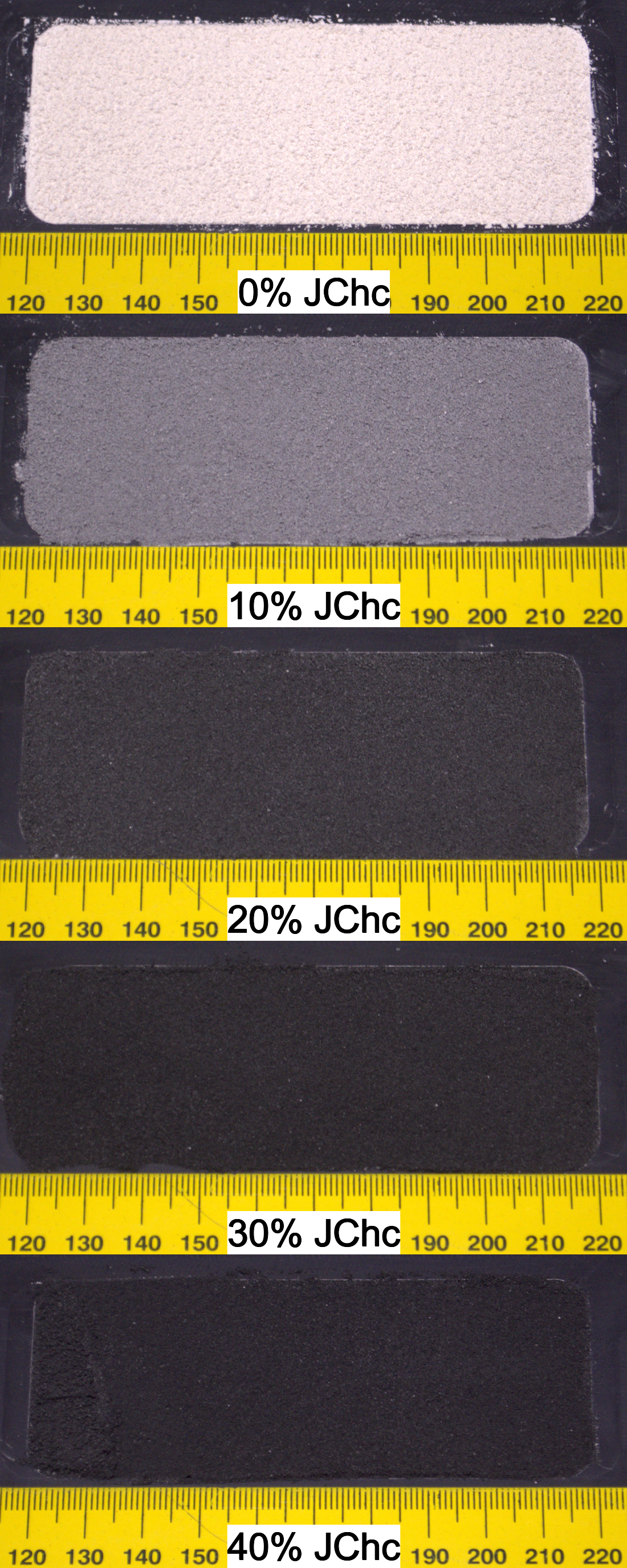}
   \end{minipage}\hfill
   \begin{minipage}[r]{0.49\linewidth}
   \includegraphics[width=\linewidth]{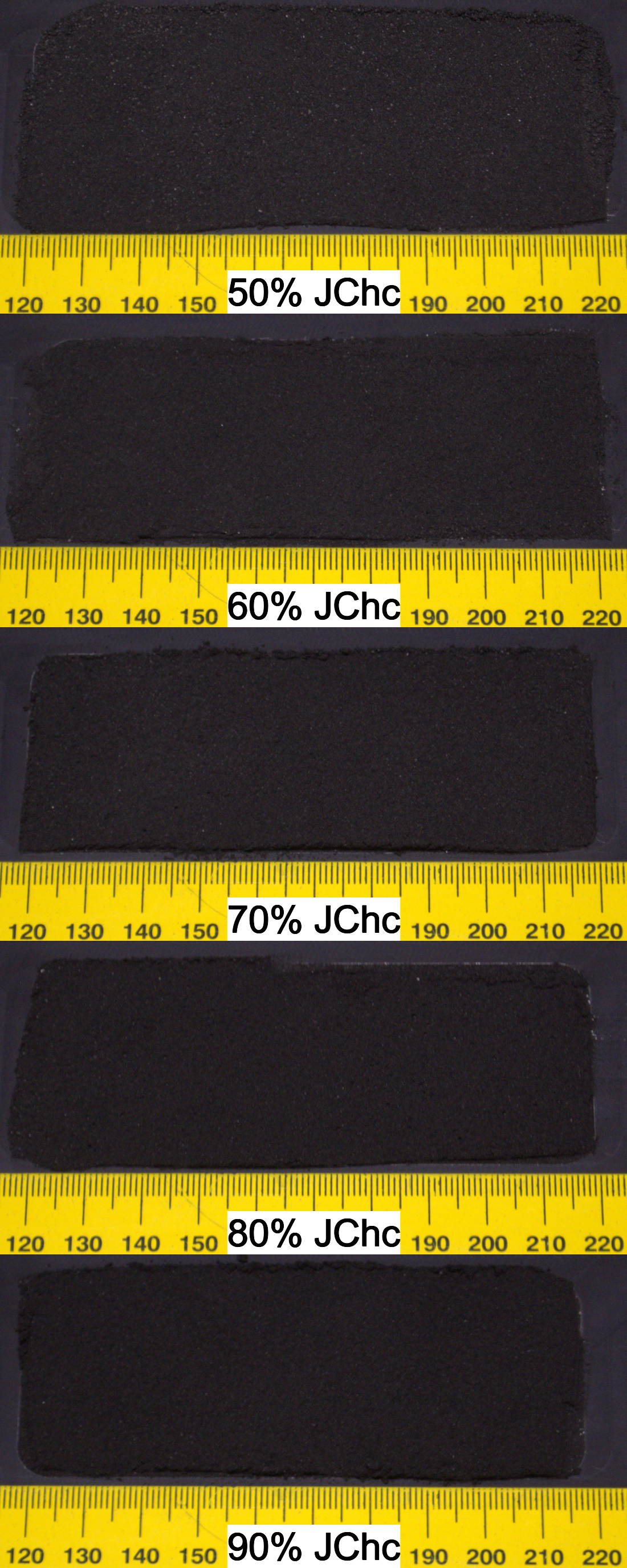}
   \end{minipage}
  \end{minipage}
  \begin{minipage}[b]{0.90\linewidth}
   \begin{minipage}[l]{0.49\linewidth}
   \caption{\label{fig_supp:pics} Pictures of the investigated samples 
            taken under the same conditions.
            These pictures were acquired with a Nikon D5100 
            camera (f/5.6, 1/50 sec, 55mm of focal, ISO 100) under 
            an emergence angle of $\sim$ 30\odeg and a main light 
            source (a white neon) placed 2 metres right above the 
            sample holder.
            }
   \end{minipage}\hfill
   \begin{minipage}[r]{0.49\linewidth}
   \includegraphics[width=\linewidth]{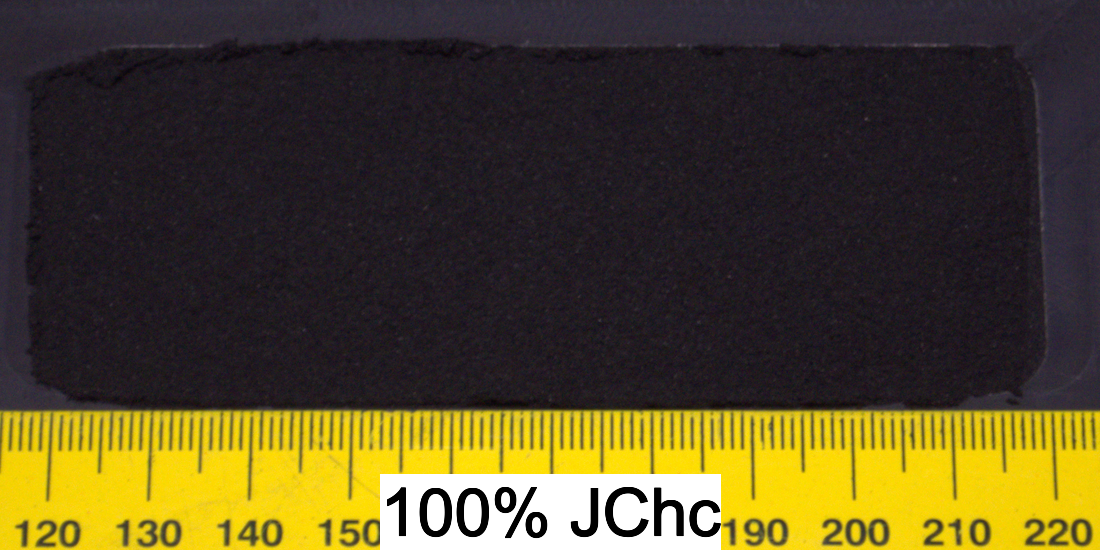}
   \end{minipage}

  \end{minipage}
 \end{center}
 \end{figure*}
\FloatBarrier

\subsection{Plots of the samples' phase curves and quality-fits}
 \label{app:phot}
 \begin{figure*}
 \begin{center}
  \begin{minipage}[t]{0.90\linewidth}
   \begin{minipage}[l]{0.49\linewidth}
   \includegraphics[width=\linewidth]{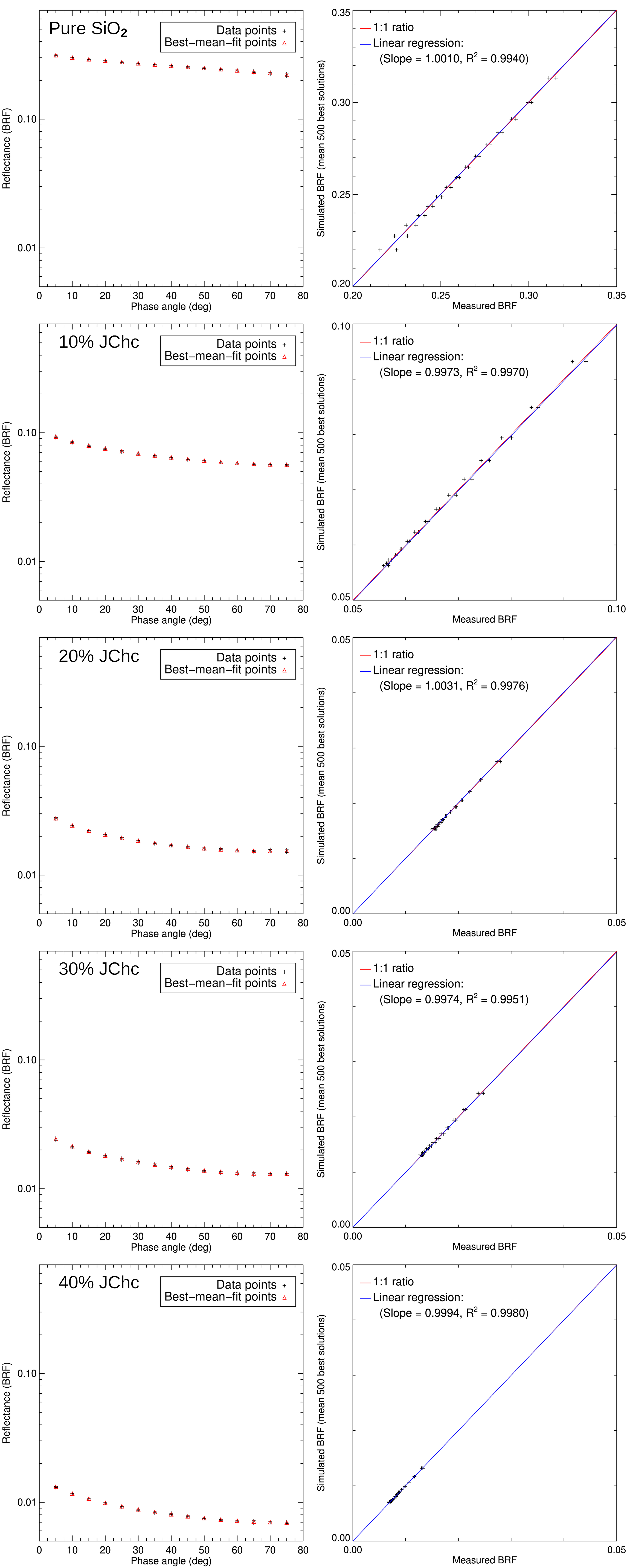}
   \end{minipage}\hfill
   \begin{minipage}[r]{0.49\linewidth}
   \includegraphics[width=\linewidth]{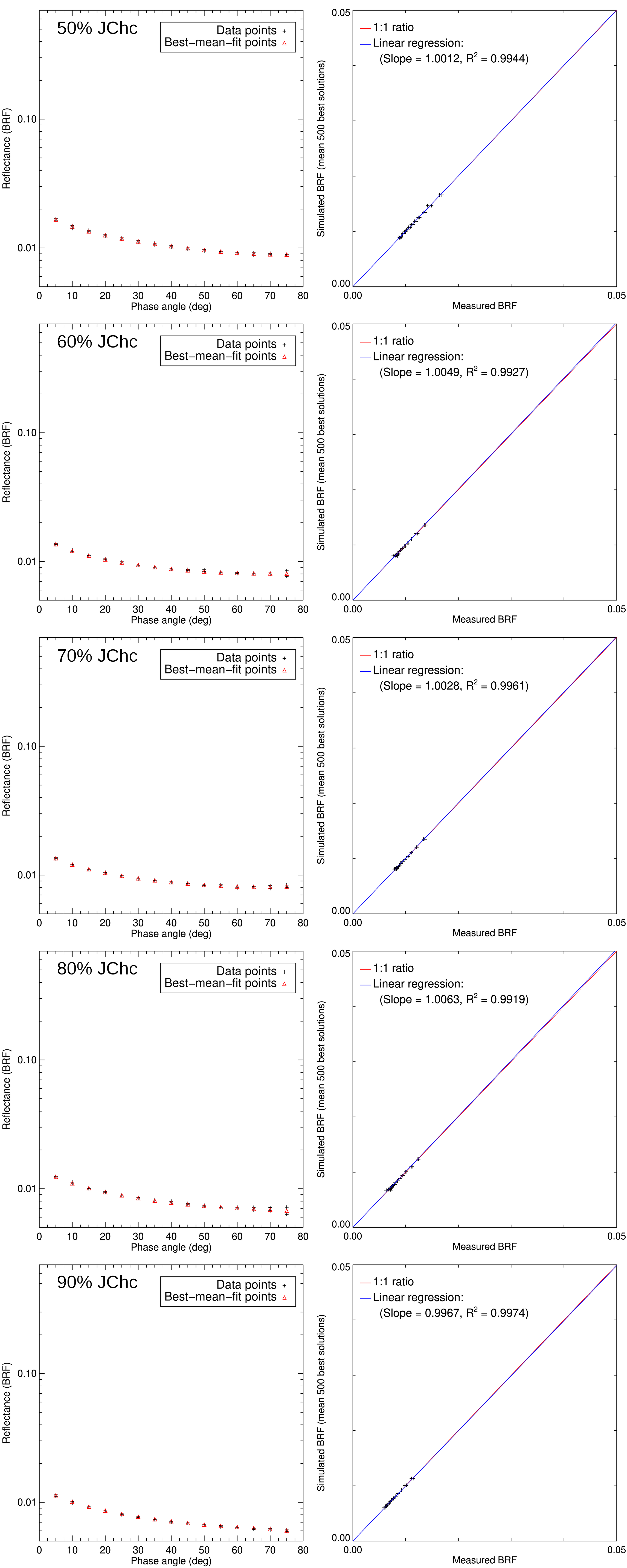}
   \end{minipage}
  \end{minipage}
  \begin{minipage}[b]{0.90\linewidth}
   \begin{minipage}[l]{0.49\linewidth}
   \includegraphics[width=\linewidth]{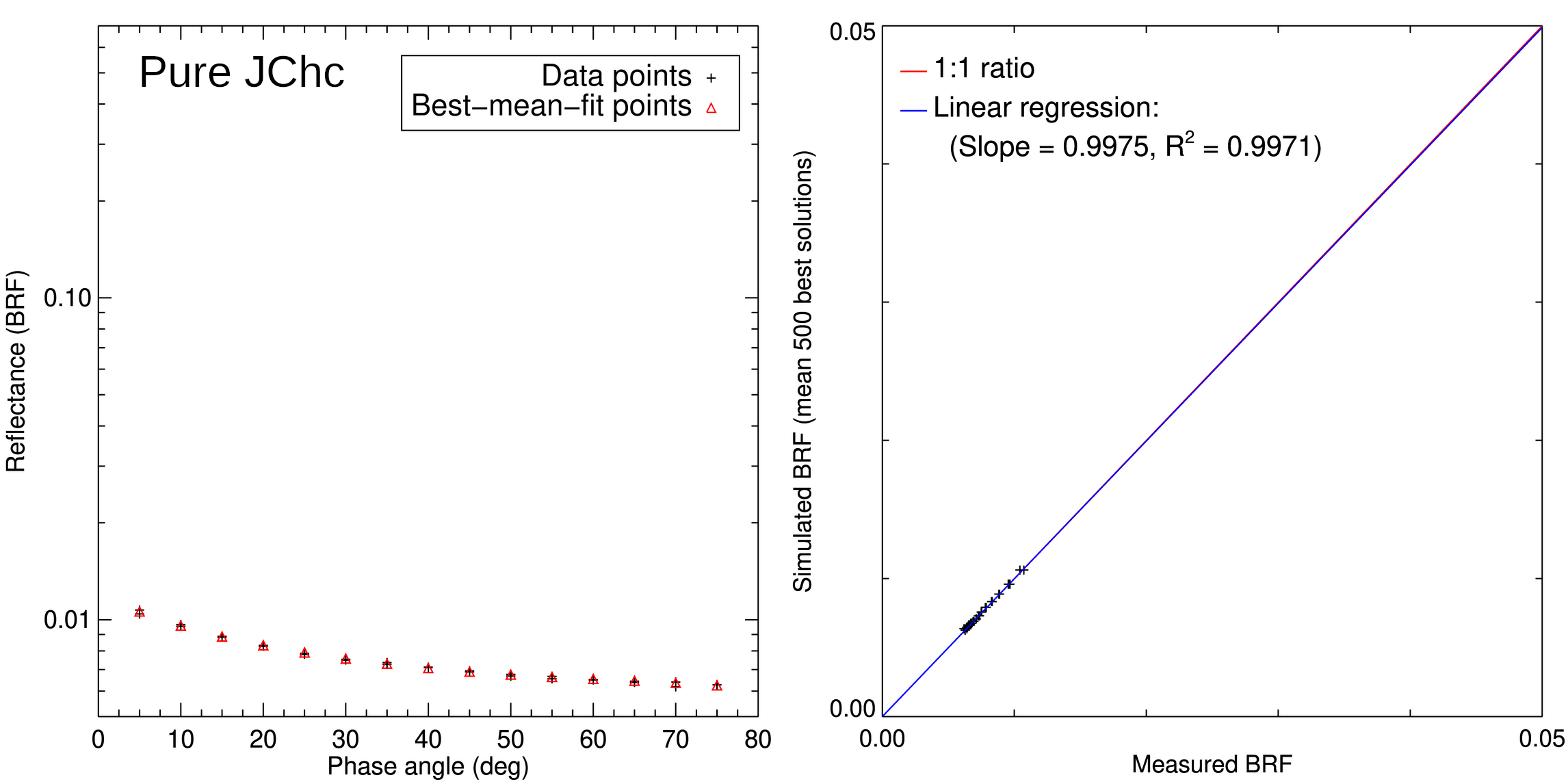}
   \end{minipage}\hfill
   \begin{minipage}[r]{0.49\linewidth}
     \caption{\label{fig:full_phase_curves} Phase curves of the investigated 
              samples overplotted with the curves of the average best-fitting 
              solutions and associated quality-fits.\\
              In each case, the curve associated the HHS parameters listed in 
              Table \ref{tab:hpk_best_sols} appropriately the measured 
              reflectance values of either end-members and all the considered 
              mixtures.}
   \end{minipage}
  \end{minipage}
 \end{center}
 \end{figure*}


\bsp	
\label{lastpage}
\end{document}